\documentclass[reprint,superscriptaddress,amsmath,amssymb,aps,preprintnumbers]{revtex4-2}
\usepackage{graphicx}
\usepackage{caption}
\usepackage{dcolumn}
\usepackage{bm}
\usepackage[usenames]{color}
\usepackage[normalem]{ulem}
\RequirePackage{graphicx,epstopdf}
\usepackage{mathpazo}
\usepackage[colorlinks=true, pdfstartview=FitV, linkcolor=blue, citecolor=blue, urlcolor=blue]{hyperref}

\graphicspath{{./figures/}}
\newcommand{\average}[1]{\ensuremath{\langle#1\rangle}}

\begin{document}
\preprint{RIKEN-iTHEMS-Report-25 
}
\title{
Achieving angular-momentum conservation with physics-informed neural networks
\\
in computational relativistic spin hydrodynamics
}
\author{Hidefumi Matsuda}
\email{da.matsu.00.bbb.kobe@gmail.com}
\affiliation{Zhejiang Institute of Modern Physics, Department of Physics, Zhejiang University, Hangzhou, 310027, China}

\author{Koichi Hattori}
\email{koichi.hattori@zju.edu.cn}
\affiliation{Zhejiang Institute of Modern Physics, Department of Physics, Zhejiang University, Hangzhou, 310027, China}
\affiliation{Research Center for Nuclear Physics (RCNP), Osaka University, Osaka 567-0047, Japan}

\author{Koichi Murase}
\email{kmurase@rcnp.osaka-u.ac.jp}
\affiliation{Research Center for Nuclear Physics (RCNP), Osaka University, Osaka 567-0047, Japan}
\affiliation{Department of Physics, Tokyo Metropolitan University, 1-1
Minami-Osawa, Hachioji, 192-0397, Tokyo, Japan}
\affiliation{RIKEN Interdisciplinary Theoretical and Mathematical Science
Program (iTHEMS), 2-1 Hirosawa, Wako, 351-0198, Saitama, Japan}

\begin{abstract}
We propose physics-informed neural networks (PINNs) as a numerical solver for relativistic spin hydrodynamics
and demonstrate that the total angular momentum, i.e., the sum of orbital and spin angular momentum, is accurately conserved throughout the fluid evolution by imposing the conservation law directly in the loss function as a training target.
This enables controlled numerical studies of the mutual conversion between spin and orbital angular momentum, a central feature of relativistic spin hydrodynamics driven by the rotational viscous effect. 
We present two physical scenarios for a rotating fluid confined in a cylindrical container: 
one case in which initial orbital angular momentum is converted into spin angular momentum in analogy with the Barnett effect, and the opposite case in which initial spin angular momentum is converted into orbital angular momentum in analogy with the Einstein-de Haas effect. 
Assuming translational invariance along the axial direction and a rotational symmetry, 
we reduce the dynamics to an effective (1+1)-dimensional problem in time and radial coordinates 
and investigate these conversion processes governed by the rotational viscous effect by analyzing 
the spacetime profiles of thermal vorticity and spin potential. Our PINNs-based framework provides the first numerical evidence for spin-orbit angular momentum conversion with fully nonlinear computational relativistic spin hydrodynamics.
\end{abstract}

\maketitle
\section{Introduction}\label{Sec:I}
Macroscopic transport of angular momentum plays a fundamental role in a wide range of physical phenomena.
For quark-gluon plasma (QGP) created by noncentral high-energy heavy-ion collisions,
intensive studies have been put on the spin polarization/alignment induced by not only a large net angular momentum~\cite{STAR:2017ckg,STAR:2018gyt,STAR:2020xbm,STAR:2021beb,ALICE:2019onw,ALICE:2021pzu} but also local vorticity~\cite{STAR:2019erd,ALICE:2019aid,STAR:2021beb,ALICE:2022dyy,STAR:2008lcm,STAR:2022fan} as proposed in early theoretical predictions~\cite{Voloshin:2004vk,Liang:2004ph,Liang:2004xn,Gao:2007bc,Becattini:2007sr,Becattini:2007nd,Becattini:2013fla,Becattini:2013vja,Becattini:2015ska,Karpenko:2016jyx}.
Controlling spin currents by mechanical rotation is a key topic in spintronics~\cite{takahashi2016spin, 
matsuo2017spin, hirohata2020review}.
Rotational motion is also essential in violent astronomical events
such as neutron star mergers~\cite{Baiotti:2016qnr}
and magnetorotational supernovae~\cite{Ardeljan:2004fq}.
Understanding these experimental observations demands
theoretical studies on analytic and computational hydrodynamics that describe
the transport of angular momentum as well as other conserved quantities in a long spacetime scale.

In particular, relativistic hydrodynamics has been a fundamental framework in the physics of QGP~\cite{Jeon:2015dfa, Denicol:2021clh, Heinz:2024jwu}, and it is important to extend it with spin degrees of freedom carried by quarks and gluons.
This framework is called relativistic spin hydrodynamics.
Stimulated by the early experimental observations of spin polarization in QGP~\cite{STAR:2017ckg,STAR:2018gyt},
rapid theoretical developments have taken place in the relativistic spin hydrodynamics~\cite{Hattori:2019lfp,Fukushima:2020ucl,Gallegos:2021bzp,Li:2020eon, Hu:2021lnx,Hu:2022azy,Singh:2022ltu,Cao:2022aku,Daher:2022wzf,Sarwar:2022yzs,
Kiamari:2023fbe,Xie:2023gbo,Ren:2024pur,Florkowski:2017ruc,    
Peng:2021ago,
Weickgenannt:2022zxs,
Weickgenannt:2023btk,Bhadury:2024ckc,
Montenegro:2017rbu,Montenegro:2020paq,
Hongo:2021ona, Hu:2021lnx,Hu:2022azy,Tiwari:2024trl,Florkowski:2024bfw, 
Fang:2024skm,Fang:2024hxa,Fang:2024sym, Wagner:2024fry,
Dey:2024cwo, She:2024rnx,Huang:2024ffg, Drogosz:2024gzv,
Wagner:2024fhf,Chiarini:2024cuv,Daher:2025pfq,
Bhadury:2025fil,Sapna:2025yss,Singh:2024cub,Singh:2025hnb,Abboud:2025shb}.
However, most studies focus on theoretical issues or analytic modeling, often relying on linearized analyses.
To step forward to phenomenological applications, it is essential to perform numerical simulations of the relativistic spin hydrodynamics.
Existing numerical studies
in Refs.~\cite{Singh:2024cub,Sapna:2025yss} have
considered the evolution of the spin degrees of freedom under a background hydrodynamic flow that is unperturbed by the spin dynamics,
but they have not yet
included the mutual conversion
between spin and orbital angular momentum.
In relativistic spin hydrodynamics, in general,
the orbital angular momentum converted from spin drives hydrodynamic flow, and spin polarization can further acquire feedback from the flow modification.
Such effects may play a role toward a resolution of the ``spin sign puzzle''~\cite{ALICE:2019aid},
which refers to a discrepancy between experimental and theoretical results in the direction of local spin polarization.
In these circumstances, it is important to establish robust and reliable simulations of relativistic spin hydrodynamics that fully incorporate
the mutual coupling between spin and flow beyond spin dynamics on top of a background flow as well as the nonlinear evolution of the fluid beyond the linearized analysis. 

\begin{figure*}[tp]
    \centering
    \includegraphics[width=1.0\textwidth]{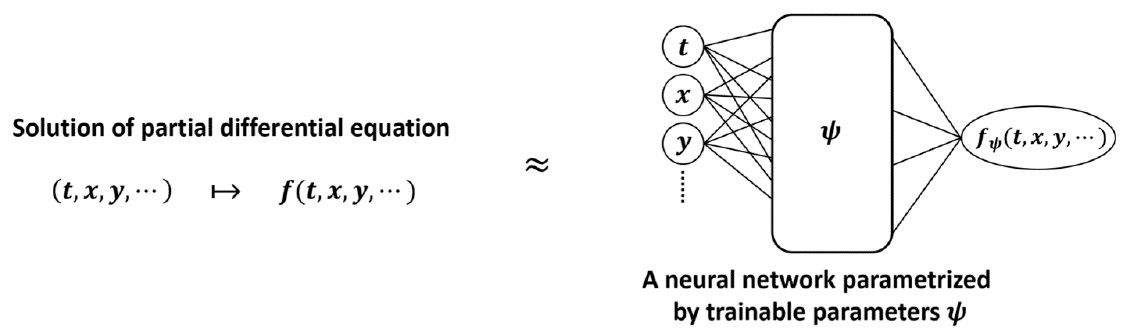}
\caption{Schematic illustration of a neural network $f_\psi$, parameterized by trainable parameters $\psi$, approximating the solution $f$ of a partial differential equation}
 \label{Fig:nn_pde}
\end{figure*}
We identify the first goal of computational spin hydrodynamics as accurately achieving the conservation of the total angular momentum, i.e., the sum of spin and orbital angular momentum.
This requirement is essential to address genuine physical consequences of the spin-orbit conversion that are free of numerical artifacts.
Regarding this point, conventional numerical frameworks, such as the family of the finite-volume method (FVM), are challenged as we discuss below.
In this work, we show that the Physics-Informed Neural Networks (PINNs) serve as a suitable method for this objective, and provide the first relativistic numerical simulation with accurate total angular momentum conservation.

PINNs have attracted attention as a {\it numerical solver} for fluid dynamics and have already been applied in numerous studies of nonrelativistic hydrodynamic equations
~\cite{raissi2020hidden,jin2021nsfnets,mao2020physics,cai2021physics}.
This application of neural networks is different from
the typical use of neural networks as an emulator of an existing simulation or experimental data.
In contrast, a neural network for PINNs takes a spacetime point, such as $(t, x, y, z)$, as the input variables and outputs field values, such as $f(t,x,y,z)$.
The neural network is trained so that the output variables satisfy the target partial differential equations supplemented by initial and boundary conditions \cite{raissi2019physics} (see Fig.~\ref{Fig:nn_pde}).
In short, the trained neural network is a representation of an approximate solution of the target differential equations, and the procedure of finding a solution is achieved by training the neural network model.
PINNs have been shown to be particularly effective for inverse problems~\cite{cai2021physics},
utilizing noisy data~\cite{eivazi2024physics},
and complex system geometries~\cite{kashefi2022physics}.
An outstanding feature of PINNs is that it provides a smooth solution as a
neural network model and is free from explicit discretization both in space and
time.  For the present purpose of the application to spin hydrodynamics, a more
important feature of PINNs is the flexibility to directly incorporate specific
physical laws of interest as training targets, which is made possible by adding
corresponding terms to the loss function.
Taking advantage of this feature, we achieve total angular momentum conservation in our numerical simulations.

\begin{figure*}[tp]
    \centering
    \includegraphics[width=0.4\textwidth]{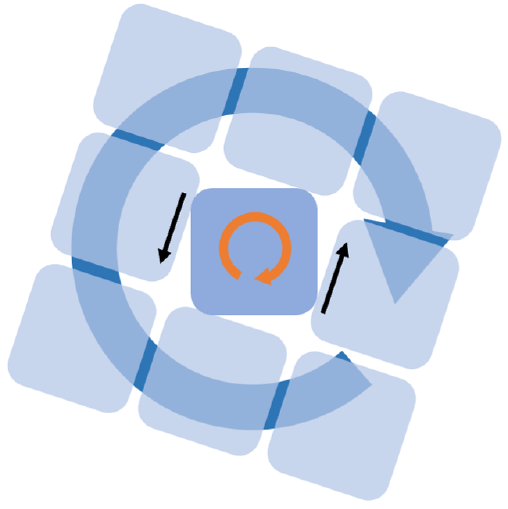}
\caption{Conceptual sketch of the rotational viscous effect}
 \label{Fig:rve}
\end{figure*}

We illustrate a rotating fluid confined in a cylindrical container.
First, we verify that our PINNs-based method can solve the spin hydrodynamic equations while accurately preserving the total angular momentum.
On top of this illustration, we investigate the {\it rotational viscous effect} that induces the spin-orbit conversion in two physical setups.
In the first setup, we show that an initial orbital fluid flow is converted to spin polarization, which
is an analogy with the Barnett effect \cite{barnett1915magnetization} and mimics the situation in QGP~\cite{Hattori:2019lfp,Fukushima:2020ucl,Gallegos:2021bzp,Li:2020eon, Hu:2021lnx,Hu:2022azy,Singh:2022ltu,Cao:2022aku,Daher:2022wzf,Sarwar:2022yzs,
Kiamari:2023fbe,Xie:2023gbo,Ren:2024pur,Florkowski:2017ruc,
Peng:2021ago,
Weickgenannt:2022zxs,
Weickgenannt:2023btk,Bhadury:2024ckc,
Montenegro:2017rbu,Montenegro:2020paq,
Hongo:2021ona, Hu:2021lnx,Hu:2022azy,Tiwari:2024trl,Florkowski:2024bfw,
Fang:2024skm,Fang:2024hxa,Fang:2024sym, Wagner:2024fry,
Dey:2024cwo, She:2024rnx,Huang:2024ffg, Drogosz:2024gzv,
Wagner:2024fhf,Chiarini:2024cuv,Daher:2025pfq,
Bhadury:2025fil,Sapna:2025yss,Singh:2025hnb,Abboud:2025shb} (see also~\cite{Weickgenannt:2020aaf,Yang:2020hri, Bhadury:2020cop,Wang:2020pej,Sheng:2021kfc,Muller:2021hpe,Kumar:2022ylt,Kumar:2023ghs,Hu:2021pwh,Hu:2022lpi,Hu:2022xjn,Hongo:2022izs,Hidaka:2023oze,Fang:2024vds,
Lin:2024cxo,Garbiso:2020puw,Gallegos:2020otk,Hashimoto:2013bna,Becattini:2009wh,Becattini:2012pp,
Becattini:2018duy,Liu:2021uhn,Becattini:2021suc,Fu:2021pok,Lin:2022tma,Lin:2024zik}).
In the second setup, we show that an initial spin polarization is converted to orbital fluid flow, which is similar to the Einstein--de Haas effect~\cite{richardson1908mechanical,einstein1915experimental}.
For both simulation setups, we find that the time evolution of the hydrodynamic flow is strongly influenced by the rotational viscous effect stemming from the coupling between spin and flow.
This is expected to further cause nonlinear time evolution of the hydrodynamic flow,
which underscores the importance of solving the fully nonlinear evolution of relativistic spin hydrodynamics. 
We also observe the rotational viscous effect as a macroscopic manifestation of the relaxation of slip between rotations of microscopic spin degrees and the ambient fluid motion
(cf. Fig.~\ref{Fig:rve}).

It is worth noting the difficulty of ensuring the angular momentum conservation
in conventional numerical schemes.  The essential part of hydrodynamic
equations is conservation laws, for which schemes based on FVM are widely used.
In FVM, the space is discretized into a finite
number of control volumes (i.e., fluid cells) and the state of the fluid is
represented as a set of conserved quantities contained in cells.  The
transport of the conserved quantities among cells are determined by fluxes
evaluated at cell interfaces, which ensures the conservation of the
quantities to be free from discretization errors.
However, it is generally nontrivial to maintain the
conservation of angular momentum simultaneously with that of energy and linear
momentum.  The reason is that FVM guarantees conservation only for the
quantities explicitly solved, whereas the conservation of the angular momentum,
following from those of the energy and linear momentum, generally receives
discretization errors.
This becomes an issue when discussing the
conversions between spin and orbital angular momentum, which can be a small
fraction of the total angular momentum.

In astrophysical contexts with the axial
symmetry, the angular momentum conservation is achieved by discretizing the
continuity equation for the angular momentum instead of that for
the linear momentum~\cite{Dimmelmeier:2002bk},
while the linear momentum conservation is trivial because of the axial symmetry.
However, in the general case, it is still difficult to ensure the conservation
of the angular and linear momenta simultaneously, which will be important in
the application of relativistic spin hydrodynamics to high-energy heavy-ion
collisions.  Simultaneous conservation of energy, and linear and angular
momenta may be achieved by extending the fluid state with new variables related
to the rotational fluid motion inside each cell and taking these into account for
the spatial reconstruction of the fluid fields and determining the
flux~\cite{despres:hal-01065105}.  Another way is to adjust the velocity field
so that all the angular momentum is carried by the center-of-mass motion of
each cell~\cite{2017JCoPh.337..289C}, neglecting the rotational motion of the
matter within the cell.  However, these directions complicate the
implementation significantly in spin hydrodynamics, where the microscopic spin
degrees also carry angular momentum.  It is also nontrivial how much the
scheme's details for preserving the orbital angular momentum leave side effects
in the spin sector.  Therefore, it is advantageous to obtain continuous
solutions with PINNs, which do not assume any explicit discretization, while
ensuring physical constraints including the angular momentum conservation.

In Sec.~\ref{Sec:2_Framework}, we present a theoretical framework of relativistic spin hydrodynamics up to the first order in the gradient expansion.
We choose a pseudo-gauge of the energy-momentum (EM) tensor such that the spin tensor becomes totally antisymmetric, and adopt the Landau frame.
In Sec.~\ref{Sec:2nd}, we derive a M\"uller-Israel-Stewart (MIS)-type relaxation equation~\cite{Mueller:1967m1, Israel:1976tn, Israel:1979wp}
for the second-order couple-stress tensor (i.e., the spatial antisymmetric part of the EM tensor), based on the entropy current analysis.
After summarizing the thermodynamic relations used in our calculations in Sec.~\ref{Sec:thermo},
we obtain the spin hydrodynamic equations of a fluid confined in a cylindrical container in Sec.~\ref{Sec:2_disk}\@.
In Sec.~\ref{Sec:3_PINNs}, we introduce a numerical framework based on PINNs for solving the spin hydrodynamic equations.
We describe the input and output variables of the neural network, as well as the construction of the loss function.
In Sec.~\ref{Sec:3_Metho},
we show that the loss function decreases and converges during the training process successfully.
In Secs.~\ref{Sec:4_o_to_s} and~\ref{Sec:4_s_to_o}, we present numerical results on the conversion between spin and orbital angular momentum,
and its inverse process, both driven by the rotational viscous effect.
In Sec.~\ref{Sec:S}, we provide a summary of this study.
In the appendices, we discuss details including a comparison between the violation of the conservation of the total angular momentum with and without an explicit constraint in the loss function.

Our conventions are as follows. The metric is denoted by $\eta_{\mu\nu}=\mathop{\mathrm{diag}}(-,+,+,+)$, and we normalize the totally antisymmetric symbol as $\epsilon^{0123}=1$.
The fluid four-velocity $u^{\mu}$ is normalized as $u^{\mu}u_{\mu}=-1$ and defines the projection operator as $\Delta^{\mu\nu} \equiv \eta^{\mu\nu} + u^{\mu}u^{\nu}$.
For shorthand notations, we define the expansion rate by $\theta \equiv \nabla_\mu u^\mu$,
the convective derivative by $D \equiv u^\mu \nabla_\mu$,
and the spatial derivative by $\nabla^\mu_\perp\equiv \Delta^{\mu\nu}\nabla_\nu$, where $\nabla_\mu$ is the covariant derivative.
Symmetrization and antisymmetrization of tensors are denoted by the following brackets, respectively:
$A^{(\mu \nu)} = (A^{\mu \nu} + A^{\nu \mu}) / 2$ and $A^{[\mu \nu]} = (A^{\mu \nu} - A^{\nu \mu}) / 2$.

\section{Relativistic Spin Hydrodynamics}\label{Sec:2}
We first provide a formulation of relativistic spin hydrodynamics, which describes the macroscopic transport of spin angular momentum,
treated as a quasi-conserved quantity, in addition to fundamental conserved quantities such as energy and momentum.
A set of differential equations that we solve in the subsequent sections is given in Eqs.~\eqref{Eq:hydro1}--\eqref{Eq:hydro5}.

\subsection{Relativistic Spin Hydrodynamics up to First Order}\label{Sec:2_Framework}
We begin with the theoretical framework of (3+1)-dimensional relativistic spin hydrodynamics up to the first order in the derivative expansion, formulated in general coordinates.

The governing equations of relativistic spin hydrodynamics are continuity equations for energy, momentum, and angular momentum,
which are the Noether charges associated with symmetries under temporal, and spatial translations and Lorentz transformations,
i.e., rotations and boosts.
The Noether currents corresponding to these charges are the EM tensor $\Theta^{\mu\nu}$
and the total angular momentum tensor $J^{\mu\nu\xi}$, whose continuity equations are expressed as
\begin{align}
\nabla_{\mu} \Theta^{\mu\nu} &= 0\ , \\
\nabla_{\mu} J^{\mu\nu\xi} &= 0\ .
\end{align}
Based on prior knowledge from the field theory, the total angular momentum tensor is assumed to be composed of orbital and spin parts as
\begin{align}
J^{\mu\nu\xi} = (x^{\nu} \Theta^{\mu\xi} - x^{\xi} \Theta^{\mu\nu}) + \Sigma^{\mu\nu\xi}\ ,
\end{align}
where $\Sigma^{\mu\nu\xi}$ denotes the spin tensor.
Then, the continuity equation for the total angular momentum tensor reads
\begin{align}
\nabla_{\mu} \Sigma^{\mu\nu\xi} = -2 \Theta^{[\nu\xi]} \ .\label{Eq:spin_ceq}
\end{align}

Note that the definitions of the EM tensor and angular momentum tensor are not unique:
one may simultaneously redefine $\Theta^{\mu\nu}$ and $\Sigma^{\mu\nu\xi}$ as
\begin{align}
\Theta^{\mu \nu} &\to \Theta'^{\mu \nu} = \Theta^{\mu \nu} - \nabla_\lambda G^{\lambda \mu \nu}\ , \\
\Sigma^{\mu \nu \xi} &\to \Sigma'^{\mu \nu \xi} = \Sigma^{\mu \nu \xi} + 2 G^{\mu [\nu \xi]}\ ,
\end{align}
while the continuity equations still hold.
Here, the tensor $G^{\mu \nu \xi}$ is antisymmetric with respect to the first two indices, $\mu$ and $\nu$.
This freedom of redefinition is referred to as a pseudo-gauge.
The choice of the pseudo-gauge determines how the total angular momentum is decomposed into spin and orbital parts.
We adopt the pseudo-gauge in which the spin tensor $\Sigma^{\mu\nu\xi}$ is totally antisymmetric with respect to all its indices.
The totally antisymmetric pseudo-gauge is consistent with the structure of the spin tensor for Dirac fermions.
In this gauge, the spin density, defined as
\begin{align}
S^{\mu\nu} = -u_\lambda \Sigma^{\lambda\mu\nu}\ ,\label{Eq:match_s}
\end{align}
satisfies the Frenkel condition,
\begin{align}
u_\mu S^{\mu\nu} = 0\ .
\end{align}
This condition means that the spin has no temporal component in the local rest frame of the fluid element,
and thus possesses only three independent degrees of freedom associated with spatial rotations in that frame.
Indeed, by projecting the continuity equation along the fluid four-velocity $u^\mu$ under the Frenkel condition, one obtains
\begin{align}
u_\nu \nabla_\mu \Sigma^{\mu\nu\xi} = -2 u_\nu \Theta^{[\nu\xi]}\ ,\label{Eq:const}
\end{align}
which represents three independent constraints.
In the local rest frame, Eq.~\eqref{Eq:const} contains no temporal derivatives, which makes it clear that the relation is nondynamical and imposes a constraint.

In the framework of spin hydrodynamics, spin is regarded as a quasi-conserved quantity
that relaxes much more slowly than other nonhydrodynamic modes.
Accordingly, the first law of thermodynamics and the Euler relation are modified so as to account for the spin density $S^{\mu \nu}$ and
its conjugate variable $\omega_{\mu \nu}$,
\begin{align}
Tds &= de     - \omega_{\mu \nu} d S^{\mu \nu}\ ,\label{Eq:thermo1}\\
T s &=  e + P - \omega_{\mu \nu} S^{\mu \nu}\ .\label{Eq:thermo2}
\end{align}
Here, $\omega_{\mu \nu}$ is referred to as the spin potential.
The spin potential $\omega^{\mu\nu}$ is antisymmetric and spatial ($u_\mu \omega^{\mu\nu}=0$), as is the spin density $S^{\mu\nu}$, reflecting its role as a Lagrange multiplier for $S^{\mu\nu}$.
The energy density is defined by projecting the EM tensor with the fluid four-velocity as
\begin{align}
e \equiv u_\mu u_\nu \Theta^{\mu\nu}\ .\label{Eq:match_e}
\end{align}
Eqs.~\eqref{Eq:match_s} and \eqref{Eq:match_e} serve as matching conditions
that relate microscopic currents of conserved quantities to thermodynamic variables in the local rest frame.
In what follows, we adopt the Landau frame,
where the fluid four-velocity is defined as
a timelike eigenvector of the symmetric part of the EM tensor,
\begin{align}
\Theta^{(\mu \nu)} u_\nu = - e u^\mu\ ,
\end{align}
with the same energy density $e$ as defined in Eq.~(\ref{Eq:match_e}).

We turn to the decomposition of the fundamental dynamical tensors, namely the EM tensor and the spin tensor,
to determine the functional dependence of these tensors on primitive variables in hydrodynamics.
We begin by presenting the general tensor decomposition of the EM tensor in the Landau frame,
\begin{align}
\Theta^{\mu\nu} = e u^\mu u^\nu + \mathcal{P}\Delta^{\mu \nu}
+\pi^{\mu\nu}+2u^{[\mu}q^{\nu]}+\phi^{\mu\nu}\ .
\end{align}
In this decomposition, the first term $e u^\mu u^\nu$ and the second term $\mathcal{P}\Delta^{\mu \nu}$
represent the temporal and spatial isotropic parts, respectively, of the energy-momentum tensor.
The third term, $\pi^{\mu\nu}$, is the spatial symmetric traceless component, referred to as the shear-stress tensor.
Furthermore, the fourth term $2u^{[\mu}q^{\nu]}$ and the fifth term $\phi^{\mu\nu}$
constitute the spatiotemporal and spatial components, respectively, of the antisymmetric part,
where $q^\mu$ is the boost heat current and $\phi^{\mu\nu}$ is the couple-stress tensor.
Note that the symmetric spatiotemporal part is excluded by the Landau condition.
We also present the general tensor decomposition of the spin tensor with a totally antisymmetric pseudo-gauge,
\begin{align}
\Sigma^{\mu\nu\xi}=
(u^\mu S^{\nu\xi}+u^\nu S^{\xi\mu}+u^\xi S^{\mu\nu})
+
X^{\mu\nu\xi}\ ,
\end{align}
where the terms in parentheses represent the temporal component, while the last term $X^{\mu\nu\xi}$ is the spatial component.

Assuming that thermodynamic variables vary slowly
in directions orthogonal to the fluid four-velocity,
fundamental dynamical tensors are approximated by a spatial gradient expansion.
At the zeroth order in the gradient expansion,
the EM tensor is expressed as
\begin{align}
\Theta^{\mu\nu}_{\rm ideal} = e u^\mu u^\nu + P \Delta^{\mu \nu}
\ ,\label{Eq:em_0}
\end{align}
where $\mathcal{P}=P$, and the terms $\pi^{\mu\nu}$, $q^\mu$, and $\phi^{\mu\nu}$ vanish.
Assuming that both the spin density and the spin potential are first-order quantities, the spin tensor has no zeroth-order component~\cite{Hattori:2019lfp,Huang:2024ffg}.
The tensor decomposition at the zeroth order describes an ideal fluid with no dissipation.
Dissipative effects specific to materials appear as higher-order corrections in the gradient expansion.
In particular, the first-order corrections are given by
\begin{align}
 \Pi & \equiv \mathcal{P}-P =-\zeta \theta\ ,\\
\pi^{\mu\nu} &= - 2\eta \sigma^{\mu\nu}\ ,\\
\phi^{\mu\nu} &= -2\gamma \rho^{\mu\nu}\ ,\label{Eq:em_1a}
\end{align}
where
\begin{align}
\sigma^{\mu \nu} &\equiv \nabla^{(\mu}_\perp u^{\nu)} - \frac{1}{3} \theta \Delta^{\mu \nu}\ , \\
\rho^{\mu \nu}   &\equiv \nabla^{[\mu}_\perp u^{\nu]} - 2 \omega^{\mu \nu}\ ,
\end{align}
and $\zeta$, $\eta$, and $\gamma$ denote the bulk, shear, and rotational viscosities, respectively~\cite{Huang:2024ffg}. 
It should be noted that the boost heat current $q^\mu$ does not appear in the first-order corrections because it is counted as a
second-order quantity in the present gradient expansion.
From the constraint equation \eqref{Eq:const}, $q^\mu$ is
expressed by a derivative of the spin tensor as 
$
q^\mu
=
-(1/2)u_\alpha\partial_\beta\Sigma^{\mu\alpha\beta}
$, which is of the second order since the spin tensor $\Sigma^{\mu\alpha\beta}$ is counted as a first-order quantity. In our order counting, the spatial part of the spin tensor, $X^{\mu\nu\xi}$,
vanishes at the first order, and the spin tensor is consequently given by
\begin{align}
\Sigma^{\mu\alpha\beta} = u^\mu S^{\alpha \beta} - u^\alpha S^{\mu \beta} + u^\beta S^{\mu \alpha}\ .\label{Eq:spin_1}
\end{align}

Substituting Eq.~\eqref{Eq:em_1a}
into Eq.~\eqref{Eq:const}, we have
\begin{eqnarray}
\nabla_\mu \Sigma^{\mu\nu\xi}
= 4 \gamma  \rho^{[\nu\xi]} \, .
\label{Eq:const-spin-orbit}
\end{eqnarray}
The divergence of the spin tensor is induced by
$\rho^{\mu\nu}$, which we call the rotation-rate mismatch.
This quantity can be rewritten as the difference between the transversely projected thermal vorticity,
$\varpi^{\mu \nu}_\perp
\equiv
\Delta^{\mu \alpha} \Delta^{\nu \beta} \partial_{[\alpha} \beta_{\beta]}$,
and the spin potential as
\begin{align}
\rho^{\mu \nu}
&=
\beta^{-1} \varpi^{\mu \nu}_\perp - 2 \omega^{\mu \nu}\ ,
\label{eq:rho}
\end{align}
where $\beta^\mu = \beta u^\mu = T^{-1} u^\mu$ is the inverse-temperature current.
Conversion between spin and orbital angular momentum occurs
when the difference in $ \rho^{\mu \nu} $ is nonzero.
Therefore, the rotational viscous effect can be interpreted as the friction
between microscopic spin degrees of freedom and the surrounding rotational
fluid motion,
as illustrated in Fig.~\ref{Fig:rve}.
The conversion rate is quantified by the rotational viscosity $ \gamma$.
In what follows, we focus on the rotational viscous effect and set the bulk and shear viscosities to zero.

\subsection{Causal spin hydrodynamics from the M\"uller-Israel-Stewart approach}\label{Sec:2nd}

The hydrodynamic equations with the zeroth- and first-order tensor decompositions given in the previous section are of parabolic type,
which implies violation of causality.
To address this issue, we incorporate higher-order corrections with the MIS approach,
extending the hydrodynamic equations to causal forms~\cite{Mueller:1967m1, Israel:1976tn, Israel:1979wp}.
Mathematically, these corrections lead to additional relaxation-type equations, as shown below,
rendering the full system of hydrodynamic equations hyperbolic and ensuring causality.
To derive these corrections, we employ the entropy current analysis.
In this analysis,
we begin by introducing an assumption for the entropy current $s^\mu$ and impose constraints on the structure of the tensor decomposition such that
its divergence remains nonnegative, $\nabla_\mu s^\mu \geq 0$, in accordance with the second law of thermodynamics.

As a starting point, we assume that the thermodynamic entropy is carried by fluid elements,
and hence the entropy current is given by $s^\mu = s u^\mu$.
This assumption will be relaxed later.
Under this assumption, using the first law of thermodynamics and the Euler relation,
the divergence of the entropy current can be written as
\begin{align}
\nabla_\mu s^\mu
=
\beta \left(De - \omega_{\mu \nu} DS^{\mu \nu}\right)
+
\beta \left( e + P - \omega_{\mu \nu} S^{\mu \nu} \right)\theta\ .
\end{align}
Within the zeroth-order tensor decomposition, shown in Eq.~\eqref{Eq:em_0},
this divergence vanishes, $\nabla_\mu s^\mu = 0$,
implying that no entropy is produced.
Then, within the first-order tensor decompositions, shown in Eqs.~\eqref{Eq:em_1a} and \eqref{Eq:spin_1},
it takes the form
\begin{align}
\nabla_\mu s^\mu = 2 \beta \gamma \rho^2 + \mathcal{O}(\nabla^3_\perp)\ .
\end{align}
From this equation, the requirement, $\nabla_\mu s^\mu \geq 0$, leads to a constraint that the rotational viscosity $\gamma$ must be a nonnegative quantity.

Here, we revisit the assumption of the entropy current, $s^\mu = s u^\mu$,
which is not justified in general.
To relax such a strong assumption,
we add a second-order term proportional to $\phi^{\alpha\beta}\phi_{\alpha\beta}$ to the entropy current $s^\mu=su^\mu$ as
\begin{align}
s^\mu = su^\mu - \frac{A}{2}\beta^\mu \phi^{\alpha\beta}\phi_{\alpha\beta}\ ,\label{Eq:s2}
\end{align}
where $A$ is an undetermined constant.
Here, we regard that the couple-stress tensor $\phi^{\mu\nu}$ contains not only the first-order contribution, $-2\gamma \rho^{\mu\nu}$ in Eq.~(\ref{Eq:em_1a}), but also some higher-order contributions.
Computing its divergence yields
\begin{align}
\nabla_\mu s^\mu
&= -\beta\phi^{\mu\nu}\Bigl[
  \rho_{\mu\nu}
  + \frac{A}{2}(\theta + \beta^{-1}D\beta)\phi_{\mu\nu}
  + AD\phi_{\mu\nu}
\Bigr]
\nonumber\\
&+ \mathcal O(\nabla^3_\perp)\ .
\end{align}
To guarantee $\nabla_\mu s^\mu \geq0$, we impose the following constitutive equation,
\begin{align}
\phi^{\mu\nu}
= -2\gamma
  \Delta^\mu{}_\alpha \Delta^\nu{}_\beta
  \Bigl[
    \rho^{\alpha\beta}
    + \frac{A}{2}(\theta + \beta^{-1}D\beta)\phi^{\alpha\beta}
    + AD\phi^{\alpha\beta}
  \Bigr]\ .
\end{align}
This equation describes the relaxation of $\phi^{\mu\nu}$ toward $-2\gamma\rho^{\mu\nu}$, which becomes manifest when rearranged as
\begin{align}
\tau_\phi
\Delta^\mu_{\ \alpha}
\Delta^\nu_{\ \beta}
D\phi^{\alpha\beta}
= - 2\gamma \rho^{\mu\nu}
  - \phi^{\mu\nu}
  - \frac{\tau_\phi}{2}\Theta \phi^{\mu\nu}\ ,\label{Eq:IS}
\end{align}
where $\tau_\phi\equiv2\gamma A$ is the relaxation time for the couple-stress tensor,
and $\Theta\equiv\theta + \beta^{-1}D\beta$.
The couple-stress tensor
$\phi^{\mu\nu}$ is now regarded as a dynamical variable, and Eq.~\eqref{Eq:IS}
serves as the equation of motion for $\phi^{\mu\nu}$.  We finally note that, in
the limit of vanishing $A$ and $\tau_\phi = 2\gamma A$ where Eq.~\eqref{Eq:IS}
reduces to the first-order case, $\phi^{\mu\nu} = -2\gamma\rho^{\mu\nu}$, the
entropy current~\eqref{Eq:s2} becomes consistent with our initial assumption
$s^\mu = su^\mu$.  Thus, the assumption $s^\mu = su^\mu$ corresponds to
assuming the vanishing relaxation time of $\phi^{\mu\nu}$ within the MIS
picture.

\begin{figure*}[tp]
    \centering
    \includegraphics[width=0.5\textwidth]{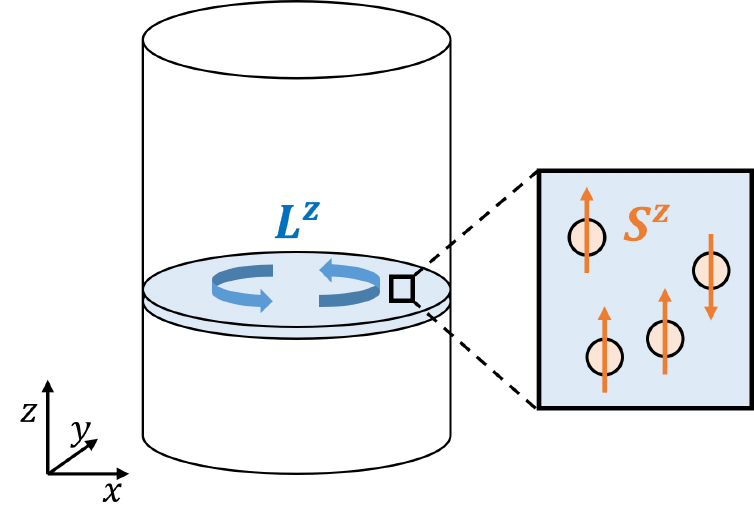}
\caption{Illustration of fluid dynamics in a 2D spatial disk}
 \label{Fig:cylinder}
\end{figure*}
\subsection{Thermodynamic relations}\label{Sec:thermo}
Finally, to make the hydrodynamic equations a closed equation system,
it is necessary to provide thermodynamic inputs, which make relations among the hydrodynamic variables and reduce the number of independent degrees of freedom.
We regard the energy density, fluid four-velocity, spin density, and couple-stress tensor as independent variables,
while the temperature, pressure, and spin potential are regarded as dependent variables,
determined through the thermodynamic relations, namely equation of state, heat capacity, and spin susceptibility.
For simplicity, we adopt the thermodynamic relations for a rarefied gas consisting of massless two-flavor fermions and gluons in thermal equilibrium.
In this setup, the equation of state takes the form
\begin{align}
P = \frac{1}{3} e = \frac{37}{90} \pi^2 T^4\ ,
\end{align}
where the factor $1/3$ reflects the conformal nature of the system.
Accordingly, the term $\beta^{-1}D\beta$ appearing in Eq.~\eqref{Eq:IS} is expressed as
\begin{align}
\beta^{-1} D\beta = - \beta C_v^{-1} D e\ ,
\end{align}
where
$C_v \equiv \frac{\partial e}{\partial T} = 4 \beta e$ is the heat capacity.
Under the same setup, the spin susceptibility is given by
\begin{align}
\chi = \frac{19}{3} \left( \frac{30e}{37\pi^2} \right)^{1/2}\ .
\end{align}
The derivation of the spin susceptibility is provided in Appendix~\ref{App:spin_sus}.
We relate the spin potential to the spin density
by assuming a linear relation between them:
\begin{align}
S^{\mu\nu} = \chi \omega^{\mu\nu}\ .
\end{align}

\subsection{Relativistic Spin Hydrodynamic Equations in a 2D Disk}\label{Sec:2_disk}
Here, we apply the hydrodynamic equations of a (3+1)-dimensional relativistic spin fluid, derived in Secs.~\ref{Sec:2_Framework}--\ref{Sec:thermo},
to a fluid confined in a cylinder in the flat spacetime.
We assume translational invariance along the axial ($z$-) direction and the absence of parity-odd tensor components with respect to $z$,
such as $u^z=0$, $\phi^{\mu z}=0$, and $S^{\mu z}=0$.
Under the latter assumption, the fluid dynamics is effectively restricted to a two-dimensional (2D) spatial disk in the transverse plane,
as shown in Fig.~\ref{Fig:cylinder}.

The set of hydrodynamic equations is written in the cylindrical coordinate system.
The rotational symmetry in the transverse plane is further assumed, simplifying the setup into a two-dimensional problem in time $t$ and radial direction $r$.
In traditional computational fluid dynamics based on FVM, it is required to
choose the density of conserved quantities, such as $\Theta^{0\mu}$ and
$J^{0\alpha\beta}$, as state variables of the fluid.  In contrast, PINNs do not
have this restriction.  Instead of the conserved quantities, we choose the
primitive variables $(e,u^\mu,\Sigma^{\mu\alpha\beta})$ as the
state variables to avoid the complexity of recovering the primitive variables
from the conserved quantities.
Taken together, we have five independent hydrodynamic variables: $e$, $u^r$, $u^\varphi$, $S^z \equiv \Sigma^{txy}$ and $\phi^{r\varphi}$,
and their evolution equations are given as follows.
\begin{itemize}
  \item Energy conservation ($u_\nu \nabla_\mu \Theta^{\mu\nu}=0$):
\begin{align}
De = - \frac{4}{3} e \theta + u_\nu \partial_\mu \phi^{\mu\nu}\ .\label{Eq:hydro1}
\end{align}
  \item Momentum conservation in the radial and azimuthal directions
  ($\Delta^{\lambda}_{\ \nu} \nabla_\mu \Theta^{\mu\nu} = 0\ \text{for } \lambda = r, \varphi$):
\begin{align}
\frac{4}{3}e D u^{ r    } &= - \frac{1}{3}u^{ r    } D e - \frac{1}{3}\nabla^{ r    } e - \Delta^{ r    }_\nu \nabla_\mu \phi^{\mu \nu}\ ,\label{Eq:hydro2}\\
\frac{4}{3}e D u^{\varphi} &= - \frac{1}{3}u^{\varphi} D e - \Delta^{\varphi}_\nu \nabla_\mu \phi^{\mu \nu}\ .\label{Eq:hydro3}
\end{align}
  \item Conservation of total angular momentum along the $z$ axis ($\nabla_\lambda \Sigma^{\lambda xy} = -2 \Theta^{[xy]}$):
\begin{align}
  \partial_t S^z &= -2 r \phi^{r \varphi}\ .\label{Eq:hydro4}
\end{align}
  \item Relaxation equation for the couple-stress tensor:
\begin{align}
\tau_\phi
\Delta^r_{\ \alpha}
\Delta^\varphi_{\ \beta}
D\phi^{\alpha\beta}
  =- 2\gamma \rho^{r \varphi}
  - \phi^{r \varphi}
  - \frac{2\tau_\phi}{3} \theta \phi^{r \varphi}\ .\label{Eq:hydro5}
\end{align}
\end{itemize}
In Eq.~\eqref{Eq:hydro5}, we make use of the energy conservation equation of a conformal ideal fluid,
$D e = -\frac{4}{3} e \theta$, for simplicity,
resulting in $\Theta=\frac{4}{3} \theta$.
The more detailed expressions of these equations are provided in Appendix~\ref{App:hydro_eq}.
The boundary and initial conditions are given in Sec.~\ref{Sec:4}, and further details of the boundary conditions are provided in Appendix~\ref{App:BC}.

\section{Numerical Methodology: PINN\lowercase{s}}\label{Sec:3}
We find solutions for relativistic spin hydrodynamics using the framework of the Physics-Informed Neural Networks (PINNs).
The PINNs framework~\cite{raissi2019physics} is based on neural networks trained with the loss function that directly incorporates differential equations, as well as initial and boundary conditions.
We employ a multilayer perceptron (MLP)
as the underlying architecture of our PINNs implementation.

To begin, we recapitulate the concept of a standard MLP before going into PINNs.
This type of neural networks is assembled from an input layer, one or more hidden layers, and an output layer. Each layer is composed of multiple interconnected units called neurons.
Then, the neural network works as a function that returns output values at the output layer when input variables are given at the input layer. When each neuron receives inputs from the neurons in the previous layer, it
applies a linear transformation followed by a nonlinear transformation using an activation function such as the hyperbolic tangent,
rectified linear unit, or sigmoid-weighted linear unit, and passes the result to the neurons in the next layer.

The universal approximation theorem~\cite{Cybenko:1989iql,Hornik:1989yye}
states that a neural network can approximate any continuous function to
arbitrary accuracy, provided it has a sufficient number of neurons.
To approximate the target output values (i.e., the desired outputs), the parameters of the network are iteratively adjusted by minimizing a loss function that quantifies the errors between the predicted and target outputs.
This iterative optimization process is known as training.
During training, the backpropagation algorithm is used to compute gradients of the loss function with respect to each parameter.
The parameters are then incrementally updated using an optimization method such as stochastic gradient descent , Adam, or RMSProp,
based on the computed gradients.

\subsection{PINNs as Hydrodynamics Solver }\label{Sec:3_PINNs}

In the framework of PINNs, the target outputs are identified with the values of
an exact solution of differential equations on randomly chosen points called
the collocation points,
so that the predicted outputs becomes close to the exact solution after a successful training (cf.~Fig.~\ref{Fig:nn_pde}).
However, the specification of the target outputs is indirect in the PINNs
framework.  Instead of directly specifying the exact solution as target
outputs, the loss function is given by the differential equations representing
the physics laws of the system.  This strategy enables us to train the model by
informing only ``the physics'' (i.e., the differential equations) without
knowing the exact solution in advance.
Complicated initial and boundary conditions may also be imposed on the
predicted outputs by incorporating them into the loss function.
As a crucial extension, the loss function can be designed to incorporate additional physical constraints that the solution must inherently satisfy, e.g., conservation laws.
We achieve the total angular momentum conservation by using this flexibility.

\subsubsection{Designing output variables}

We investigate a solution for the hydrodynamic equations given in Eqs.~\eqref{Eq:hydro1}--\eqref{Eq:hydro5},
which is a set of real-valued functions of the temporal and radial coordinates, $t$ and $r$,
defined over the time interval $t \in [0, t_{\text{max}}]$ and a cylinder of radius $R$, $r \in [0, R]$.
The input layer of the neural network consists of two neurons, while the output layer consists of five neurons.
The temporal and radial coordinates $(t, r)$ constitute input variables.
The output variables are parameterized by trainable parameters $\psi$ and consist of five components corresponding to the five hydrodynamic degrees of freedom, denoted as $\mathrm{NN}_{\psi,i}(t, r)\ (i=1,2,\ldots,5)$.

Appropriate choices of the output variables enhance the efficiency of the solution search.
To impose a part of the initial and boundary conditions efficiently, we adopt the following representation of the hydrodynamic fields:
\begin{subequations}
\label{eq:ansatz}
\begin{align}
e_\psi(t, r)              = e(t=0, r)              + r  \left[\mathrm{NN}_{\psi,1}(t, r)-\mathrm{NN}_{\psi,1}(t=0, r)      \right]&\ ,\label{Eq:ansatz1}\\
u^r_\psi(t, r)            = u^r(t=0, r)            + r  \left[\mathrm{NN}_{\psi,2}(t, r)-\mathrm{NN}_{\psi,2}(t=0, r)      \right]&\nonumber\\
                                                   - r  \left[\mathrm{NN}_{\psi,2}(t, r=R)-\mathrm{NN}_{\psi,2}(t=0, r=R)  \right]&\ ,\label{Eq:ansatz2}\\
u^\varphi_\psi(t, r)       = u^\varphi(t=0, r)       + r  \left[\mathrm{NN}_{\psi,3}(t, r)-\mathrm{NN}_{\psi,3}(t=0, r)      \right]&\ ,\label{Eq:ansatz3}\\
S^z_\psi(t, r)            = S^z(t=0, r)            + r^2\left[\mathrm{NN}_{\psi,4}(t, r)-\mathrm{NN}_{\psi,4}(t=0, r)      \right]&\nonumber\\
                                                   - r^2\left[\mathrm{NN}_{\psi,4}(t, r=R)-\mathrm{NN}_{\psi,4}(t=0, r=R)  \right]&\ ,\label{Eq:ansatz4}\\
\phi^{r\varphi}_\psi(t, r) = \phi^{r\varphi}(t=0, r) + r  \left[\mathrm{NN}_{\psi,5}(t, r)-\mathrm{NN}_{\psi,5}(t=0, r)      \right]&\nonumber\\
                                                   - r  \left[\mathrm{NN}_{\psi,5}(t, r=R)-\mathrm{NN}_{\psi,5}(t=0, r=R)  \right]&\ .\label{Eq:ansatz5}
\end{align}
\end{subequations}
By subtracting the output variable $\mathrm{NN}_{\psi,i}(t,r)$ from its values at the initial time or at the outer boundary of the cylinder, the parametrization is designed to automatically satisfy the initial conditions of $e$, $u^r$, $u^\varphi$, $S^z$, and $\phi^{r\varphi}$ at $t = 0$,
as well as the Dirichlet boundary conditions of $u^r$, $S^z$, and $\phi^{r\varphi}$ at $r = R$.
Multiplying the output variables by $r$ ensures the Dirichlet boundary conditions for $e$, $u^r$, $u^\varphi$, $S^z$, and $\phi^{r\varphi}$ at $r = 0$, while maintaining the regularity of the solution at the origin by eliminating divergences arising from the $1/r$ term in the hydrodynamic equations.
The multiplier $r^2$ in Eq.~\eqref{Eq:ansatz4} is motivated by the behavior $S^z \propto r^2$ as $r \to 0$,
which is satisfied when imposing $\phi^{r\varphi} \propto r$ as $r \to 0$, as clearly seen in Eq.~\eqref{Eq:hydro4}.
We also impose the Neumann boundary conditions on $u^r$ and $\phi^{r\varphi}$ at $r = R$, which are, however, implemented through the loss function as explained below.
Further details on the boundary conditions are provided in Appendix~\ref{App:BC}.

In Eq.~(\ref{eq:ansatz}) and followings, we attach a subscript $\psi$ to a function $f$ of the hydrodynamic variables, denoted as $f_\psi$, when the function $f$ is approximated by the output variables.
If a function $f$ also depends on temporal and spatial derivatives of the hydrodynamic variables,
it is understood that
$f_\psi$ is evaluated with automatic differentiation of the output variables.

\subsubsection{Designing the loss function}
The loss function in our calculations comprises multiple components,
each of which quantifies how well the governing differential equations,
boundary conditions, and conservation laws are satisfied during training.
The corresponding residuals are denoted as
$R^{\text{G.E.}}_i$, $R^{\text{B.C.}}_i$, and $R^{\text{C.L.}}_i$ with the superscripts for these three loss categories
and the subscript $i$ referring to individual residuals classified in each category.

In designing the loss function,
we use the approach proposed in Ref.~\cite{kendall2018multi}.
A key advantage of this approach is its ability to automatically rebalance the contributions of residuals with large-scale separations.
Without such rebalancing, there is a risk of imbalanced optimization, where the training process may overly focus on particular losses that can be reduced more easily than others.
Each residual is assumed to follow a Gaussian distribution with a trainable variance parameter, and the overall loss is defined as the sum of the corresponding negative log-likelihoods:
\begin{align}
L(\psi, \sigma) &=
    \sum_i \left[ \frac{1}{2(\sigma^{\text{G.E.}}_i)^2} \bar{R}^{\text{G.E.}}_i(\psi) + \ln \sigma^{\text{G.E.}}_i \right] \nonumber\\
& + \sum_i \left[ \frac{1}{2(\sigma^{\text{B.C.}}_i)^2} \bar{R}^{\text{B.C.}}_i(\psi) + \ln \sigma^{\text{B.C.}}_i \right] \nonumber\\
& + \sum_i \left[ \frac{1}{2(\sigma^{\text{C.L.}}_i)^2} \bar{R}^{\text{C.L.}}_i(\psi) + \ln \sigma^{\text{C.L.}}_i \right] \ ,\label{Eq:loss}
\end{align}
where $\bar{R}^{\text{G.E.}}_i$, $\bar{R}^{\text{B.C.}}_i$, and $\bar{R}^{\text{C.L.}}_i$
are the averaged residuals specified below, and $\sigma^{\text{G.E.}}_i$, $\sigma^{\text{B.C.}}_i$, and $\sigma^{\text{C.L.}}_i$
are the trainable variances.
Trained with these variance parameters, the network can automatically adjust the relative balance of residuals even when there are large separations in the magnitudes of the residuals.
This approach corresponds to introducing a likelihood model in a Bayesian framework, explicitly accounting for aleatoric uncertainty, as originally proposed in Ref.~\cite{kendall2017uncertainties}.
It was later shown in Ref.~\cite{kendall2018multi} that this formulation provides an additional benefit in multitask learning, enabling automatic balancing among different tasks.

However, even with the above balancing mechanism, we still observed imbalanced optimization. During training, the model tends to overemphasize losses from the boundary conditions and conservation laws, which can be reduced more easily than that from the governing equations. As a result, the residuals associated with the governing equations tend to be underweighted in comparison.
To address this issue, we regularize the coefficients in front of residuals in Eq.~(\ref{Eq:loss}) by introducing an upper bound on the scale differences among multiple
components.
Specifically, for each index $i$, $(\sigma^{\text{B.C.}}_i)^2$ and $(\sigma^\text{C.L.}_i)^2$ are rescaled as
\begin{align}
(\sigma^\text{B.C.}_i)^2
&\to
\max\left\{
(\sigma^\text{B.C.}_i)^2,\ 
0.01 \cdot \max_j(\sigma^\text{G.E.}_j)^2
\right\}\ ,\\
(\sigma^\text{C.L.}_i)^2
&\to
\max\left\{
(\sigma^\text{C.L.}_i)^2,\ 
0.01 \cdot \max_j(\sigma^\text{G.E.}_j)^2
\right\}\ ,
\end{align}
respectively.

Now, we are in position to provide an explicit form of each residual.
First, the averaged residual $\bar{R}^{\text{G.E.}}_i(\psi)$ is designed to measure how well the $i$th hydrodynamic equation is satisfied,
\begin{align}
\bar{R}^\text{G.E.}_i(\psi) = \mathbb{E}_{(t,r)} \left[r R^\text{G.E.}_i(\psi; t, r)^2 \right]\ ,\label{Eq:res_ge}
\end{align}
where $r$ accounts for the cylindrical volume element.
The expectation $\mathbb{E}_{(t,r)}$ denotes the average over the domain of $(t, r)$.
The explicit forms of the residuals $R^{\text{G.E.}}_i(\psi; t, r)$ follow from the hydrodynamic equations~\eqref{Eq:hydro1}--\eqref{Eq:hydro5} as
\begin{subequations}
\begin{align}
R^\text{G.E.}_1(\psi; t, r)&=u_{\psi\,\nu} \nabla_\mu \Theta^{\mu\nu}_\psi\ ,\label{Eq:res_ge1}\\
R^\text{G.E.}_2(\psi; t, r)&=\Delta^{r}_{\psi\ \nu} \nabla_\mu \Theta^{\mu\nu}_\psi\ ,\label{Eq:res_ge2}\\
R^\text{G.E.}_3(\psi; t, r)&=\Delta^{\varphi}_{\psi\ \nu} \nabla_\mu \Theta^{\mu\nu}_\psi\ ,\label{Eq:res_ge3}\\
R^\text{G.E.}_4(\psi; t, r)&=\nabla_\lambda \Sigma^{\lambda xy}_\psi + 2 \Theta^{[xy]}_\psi\ ,\label{Eq:res_ge4}\\
R^\text{G.E.}_5(\psi; t, r)&=\tau_\phi
\Delta^\mu_{\psi\ \alpha}
\Delta^\nu_{\psi\ \beta}
D_\psi \phi^{\alpha\beta}_\psi
  + 2\gamma \rho^{\mu\nu}_\psi
  + \phi^{\mu\nu}_\psi
  + \frac{\tau_\phi}{2}\Theta_\psi \phi^{\mu\nu}_\psi\ .\label{Eq:res_ge5}
\end{align}
\end{subequations}

Next, the averaged residual $\bar{R}^{\text{B.C.}}_i(\psi)$ is designed to impose the Neumann-type boundary conditions on the surface of cylinder at $r = R$, which are
given by
\begin{align}
\bar{R}^\text{B.C.}_i(\psi) = \mathbb{E}_{(t)} \left[ R^\text{B.C.}_i(\psi; t)^2 \right]\ ,
\end{align}
where
\begin{subequations}
\begin{align}
R^\text{B.C.}_1(\psi; t) &= \partial_r u^r_\psi(t,R)
-
\partial_r u^r_\psi(0,R), \\
R^\text{B.C.}_2(\psi; t) &= \partial_r \phi^{r\varphi}_\psi(t,R)
-
\partial_r \phi^{r\varphi}_\psi(0,R)
\ .
\end{align}
\end{subequations}
The expectation $\mathbb{E}_{(t)}$ denotes the average over the domain of $t$.

Finally, the averaged residuals $\bar{R}^\text{C.L.}_1(\psi)$ and $\bar{R}^\text{C.L.}_2(\psi)$ are designed to enforce the conservation of angular momentum.
The former accounts for the local conservation defined by the fulfillment of the continuity equation, while the latter the global conservation defined as the time invariance of the spatially integrated angular momentum.
Explicit forms of these residuals are given by
\begin{subequations}
\begin{align}
\bar{R}^\text{C.L.}_1(\psi) &= \mathbb{E}_{(t, r)} \left[ r R_{\text{local }}(\psi; t, r)^2 \right]\ ,\label{Eq:residual_local} \\
\bar{R}^\text{C.L.}_2(\psi) &= \mathbb{E}_{(t)}         \left[   R_{\text{global}}(\psi; t   )^2 \right]\ ,\label{Eq:residual_global}
\end{align}
\end{subequations}
where $R_{\text{local}}(\psi; t, r)$ follows from the continuity equation for the $xy$-component of the angular momentum:
\begin{align}
R_{\text{local }}(\psi; t, r) = \nabla_\lambda J^{\lambda xy}_\psi\ . \label{Eq:CL_local}
\end{align}
The other residual $R_{\text{global}}(\psi; t)$ quantifies the deviation from the angular momentum over time and is defined as\begin{align}
R_{\text{global}}(\psi; t) =
\frac{ \int dr\, r J^{txy}_\psi(t, r) - \int dr\, r J^{txy}_\psi( 0 , r) }{ \int dr\,  r J^{txy}_\psi(0, r) }\ .\label{Eq:global}
\end{align}
Note that both Eqs.~\eqref{Eq:hydro4} and~\eqref{Eq:CL_local} represent the same angular momentum conservation law, $\nabla_\lambda J^{\lambda xy} = 0$.
The former makes use of the energy-momentum conservation, $\nabla_\mu \Theta^{\mu\nu} = 0$, to rewrite the relation as the evolution equation for the spin tensor, $\nabla_\mu \Sigma^{\mu xy} = -2\Theta^{[xy]}$, whereas the latter keeps the expression without such a reduction, thereby directly
quantifying numerical violations of angular momentum conservation.

Before proceeding to simulation results, we specify how we evaluate the expectations with the discrete simulation data at the collocation points in the $t$-$r$ plane.
First, for the evaluation of
$\bar{R}^{\text{G.E.}}_i(\psi)$ and $\bar{R}^{\text{C.L.}}_1(\psi)$,
we use collocation points $\{t^{(n)}, r^{(n)}\}_{n=1}^{N_{\text{col}}}$,
sampled from a uniform distribution over $t = [0, t_{\text{max}}]$ and $r = [0, R]$,
and approximate the expectations as
\begin{align}
&\mathbb{E}_{(t, r)} \left[ r R^\text{G.E.}_i(\psi; t, r)^2 \right]\nonumber\\
&\approx
\frac{1}{N_\text{col}} \sum_{n=1}^{N_\text{col}}
r^{(n)} \left[ R^\text{G.E.}_i(\psi; t^{(n)}, r^{(n)})^2 \right]\ ,\\
&\mathbb{E}_{(t,r)} \left[r R_{\text{local }}(\psi; t, r)^2 \right]\nonumber\\
&\approx
\frac{1}{N_\text{col}} \sum_{n=1}^{N_\text{col}}
r^{(n)} \left[ R_{\text{local}}(\psi; t^{(n)}, r^{(n)})^2 \right]\ .
\end{align}
Next, for the evaluation of $\bar{R}^{\text{B.C.}}_i(\psi)$,
we use the collocation points $\{t^{(n)}, R\}_{n=1}^{N_{\text{col}}}$,
whose time coordinates are the same as those introduced above,
and approximate the expectation as
\begin{align}
&\mathbb{E}_{(t)} \left[ R^\text{B.C.}_i(\psi; t)^2 \right]
\approx
\frac{1}{N_\text{col}} \sum_{n=1}^{N_\text{col}}
\left[ R^\text{B.C.}_i(\psi; t^{(n)})^2 \right]\ .
\end{align}

Finally, for the evaluation of $\bar{R}^{\text{C.L.}}_2(\psi)$,
we use $N_{\text{col,global}} = N_t \times N_r$ collocation points,
which are Cartesian product of $\{t^{(n)}_\text{global}\}_{n=1}^{N_t}$ and $\{r^{(n)}_\text{global}\}_{n=1}^{N_r}$.
The time points $\{t^{(n)}_\text{global}\}_{n=1}^{N_t}$ are constructed as follows: the interval $[0,t_{\max}]$ is divided into $N_t$ blocks, and one point is sampled uniformly from each block, except for the last block.
The point in the last block is fixed at $t_\text{max}$ because it turned out to improve the training accuracy near the boundary.
The radial points $\{r^{(n)}_\text{global}\}_{n=1}^{N_r}$ are a uniformly spaced grid, defined as
$r^{(n)}_\text{global} = R ( n - 1/2 )/N_r$ with $n=1,2,\cdots, N_r$.
Using these points, the integral appearing in the definition of $R_{\text{global}}(\psi; t)$,
shown in Eq.~\eqref{Eq:global}, is discretized at each time slice $t^{(n)}$ as
\begin{align}
\int_0^R dr\,  r J^{txy}_\psi(t^{(n)}_\text{global}, r)
\approx
\frac{R}{N_r} \sum_{m=1}^{N_r} r^{(m)}_\text{global}
J^{txy}_\psi(t^{(n)}_\text{global}, r^{(m)}_\text{global})\ .
\end{align}
The integrals at $t=0$ in $R_{\text{global}}(\psi;t)$~\eqref{Eq:global} are evaluated using the same radial grid.
This discretization enables us to evaluate the function $R_{\text{global}}(\psi; t^{(n)})$.
Then,
$\bar{R}^{\text{C.L.}}_2(\psi)$ is approximated by averaged $R_{\text{global}}(\psi; t^{(n)})$ over $N_t$ time points as
\begin{align}
\mathbb{E}_{(t)}\left[ R_{\text{global}}(\psi; t   )^2 \right]
\approx
\frac{1}{N_t} \sum_{n=1}^{N_t}
\left[ R_{\text{global}}(\psi; t^{(n)})^2 \right]\ .
\end{align}

\subsection{Evaluation of Training Progress}\label{Sec:3_Metho}

Here, we analyze both the training convergence and the accuracy of the total angular momentum conservation.
We first summarize the numerical setup.
The number of hidden layers is 4, with 250 units in each layer.
We use the hyperbolic tangent function as the activation function.
The weights connecting the units are initialized using the Xavier method.
The loss function is minimized using the Adam optimization method, with the initial learning rate set to $10^{-3}$.
After the overall loss has sufficiently decreased, the learning rate is gradually reduced to $10^{-4}$ and then to $10^{-5}$,
thereby promoting convergence to the optimal solution.
The autograd engine is implemented using PyTorch.
The full batch size is
$2N_\text{col} + N_{\text{col,global}} = 2N_\text{col} + N_tN_r = 2 \times 50,000 + 10 \times 1,000 = 110,000$,
which corresponds to the total number of collocation points.
In the actual optimization, a set of $N_\text{col}$ collocation points are divided into mini-batches of 5,000 points.
For an update with each mini-batch,
$\bar{R}^\text{G.E.}_i(\psi)$, $\bar{R}^\text{C.L.}_1(\psi)$, and $\bar{R}^\text{B.C.}_i(\psi)$
are evaluated using the corresponding mini-batches,
while
$\bar{R}^\text{C.L.}_2(\psi)$ is always evaluated on the full $N_tN_r$ grid.
Therefore, the effective mini-batch size for one optimization step is
$2 \times 5,000 + 10 \times 1,000 = 20,000$. The training takes approximately 2 hours, using one NVIDIA H100 NVL graphics processing unit (GPU).

The initial conditions for the hydrodynamic variables are provided as
\begin{subequations}
\label{Eq:init_orb}
\begin{align}
e(t=0,r) &= \bar{e}\ ,\label{Eq:init_orb_1}\\
u^r(t=0,r) &= 0\ ,\label{Eq:init_orb_2}\\
u^{\varphi}(t=0,r) &= \delta_1 \cdot \sin^4 \left( \frac{\pi r}{R} \right)\ ,\label{Eq:init_orb_3}\\
S^z(t=0,r) &= 0\ ,\label{Eq:init_orb_4}\\
\phi^{r\varphi}(t=0,r) &= 0\ ,\label{Eq:init_orb_5}
\end{align}
\end{subequations}
where $\delta_1$ is a parameter controlling the magnitude of the initial angular velocity.
When $\delta_1 = 0$, the system is in a global thermal equilibrium state.
When $\delta_1 \neq 0$, the time evolution of the system starts from an out-of-equilibrium state, carrying finite orbital angular momentum while the initial spin angular momentum is vanishing.
The parameters used in the simulation are summarized in Table~\ref{Tab:para}\@.
These parameters as well as the quantities shown below are normalized with respect to $\bar{e}$.

We set the maximum time to $t_{\text{max}}=0.4$.
To enhance the efficiency of the learning process, 
we first train the network on a shorter time interval and then gradually extend the training domain in time.
We begin training with a limited time range, setting a tentative maximum time to $t'_{\text{max}}=0.2$.
Once the training has progressed sufficiently, we iteratively extend $t'_{\text{max}}$ to 
$\min\{1.2 t'_{\text{max}}, t_{\text{buf}}\}$,
where $t_{\text{buf}}=0.42$ is introduced as a small numerical buffer beyond the physical maximum time $t_{\text{max}}=0.4$.
Since the accuracy tends to deteriorate near the upper end of the training interval, we use only the results in the physical range $t \leq t_{\text{max}}$.

As the time domain is enlarged, both $N_\text{col}$ and $N_{\text{col,global}}=N_tN_r$ are updated accordingly.
Specifically, each time the time domain $[0,t'_{\text{max}}]$ is extended,
$N_\text{col}$ and $N_t$ are reset to new larger values, and
collocation points are regenerated over the new time domain,
while $N_r$ is kept fixed throughout the training.
The values of $N_\text{col}=50,000$ and $N_t=10$ quoted earlier in this section
are the values used at the end of the training, when the time domain has
reached its full extent $[0,t_\text{buf}]$.
Smaller values of $N_\text{col}$ and $N_t$ are used earlier in training, before the time domain
reaches $t_\text{buf}$.
When training continues within the same time domain without extension,
both sets of points are also regenerated at regular intervals.
The $N_t$ points are regenerated approximately every 5,000 training iterations,
whereas the $N_\text{col}$ points are regenerated approximately every 25,000 training iterations.
Numerical results shown in this section are obtained after 150,000 iterations for $\gamma=2$.

\begin{table}[htbp]
\begin{center}
\begin{tabular}{r|c}
\hline\hline
$\bar{e}$      & $1$ \\
$R$            & $1$ \\
$t_\text{max}$ & $0.4$ \\
$\gamma_\phi$  & $2$ \\
$\tau_\phi$    & $2$ \\
$\delta_1$     & $0.2$ \\
\hline\hline
\end{tabular}
\end{center}
\caption{
Parameters used in Sec.~\ref{Sec:3_Metho}, expressed in units of $\bar{e}$.
}
\label{Tab:para}
\end{table}

\begin{figure*}[tp]
\begin{minipage}{0.32\textwidth}
    \centering
    \includegraphics[width=\textwidth]{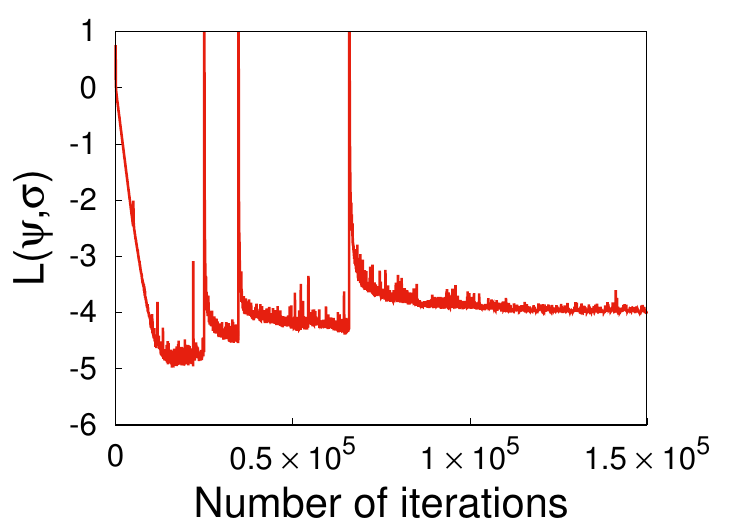}
\end{minipage}
\begin{minipage}{0.35\textwidth}
    \centering
    \includegraphics[width=\textwidth]{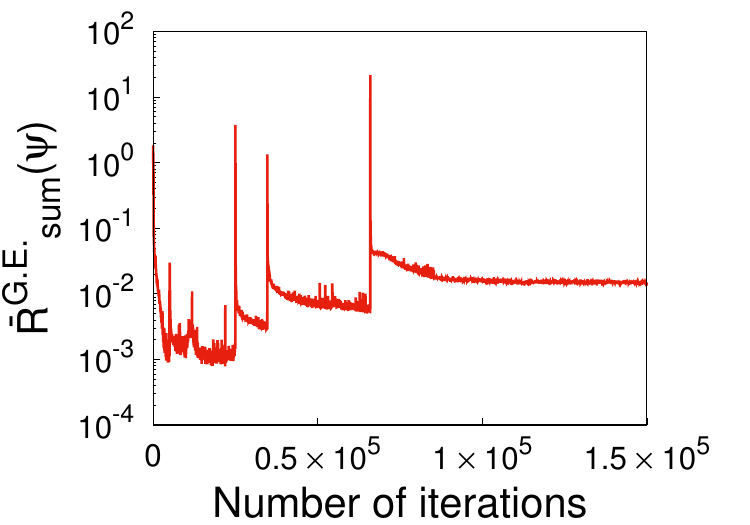}
\end{minipage}
\begin{minipage}{0.32\textwidth}
    \centering
    \includegraphics[width=\textwidth]{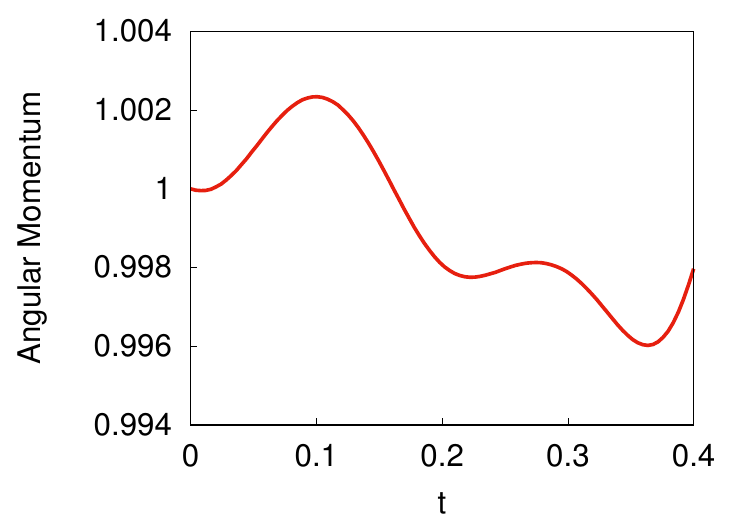}
\end{minipage}
\caption{
Left:
Decrease in the loss function \eqref{Eq:loss} as the training process proceeds.
Middle:
Decrease in the total residual of the governing equations,
$\bar{R}^\text{G.E.}_\text{sum} = \sum_i \bar{R}^\text{G.E.}_i$,
as the same training process proceeds as in the left panel.
Right:
Time evolution of the total angular momentum, accurately conserved over the 2D disk; The vertical axis is normalized by its initial value at $t=0$.
}
\label{Fig:loss_Metho}
\end{figure*}
In the left panel of Fig.~\ref{Fig:loss_Metho},
we show a decreasing behavior of the loss function \eqref{Eq:loss} as the training proceeds.
The loss function decreases rapidly in the early stage of training
after which the rate of decrease slows down and eventually reaches a plateau. This plateau signals the convergence of the training process.
The four observed jumps in the loss function are attributed to the update of the time domain mentioned above.

The middle panel of Fig.~\ref{Fig:loss_Metho} shows the decreasing behavior in the total residual of the governing equations during the same training process as in the left panel.
This total residual is computed as the sum of 
$\bar{R}^\text{G.E.}_i$ defined in Eqs.~\eqref{Eq:res_ge},
and is expressed as $\bar{R}^\text{G.E.}_\text{sum} = \sum_i \bar{R}^\text{G.E.}_i$.
It is found that $\bar{R}^\text{G.E.}_\text{sum}$ eventually falls below $1.5 \times 10^{-2}$,
although it rises tentatively on every extension of the time domain $t_{\rm max}'$,
confirming that the governing equations are well satisfied with only minor violations.

In the right panel of Fig.~\ref{Fig:loss_Metho},
we show the time evolution of the global angular momentum,
obtained by spatial integration of the angular momentum density, $J^{txy}$, over the 2D disk.
The result shown in this figure is normalized by its initial value at $t=0$.
{\it
We find a well-controlled conservation of the global angular momentum within 0.4\% of violation with respect to the initial value.
}
In Appendix~\ref{App:comp},
we also compare the cases in which the loss function includes or excludes the term that explicitly enforces the angular momentum conservation law,
and verify that the global conservation is hardly achieved without this penalty term.

Finally, we mention the conservation of net energy, defined as the spatial integral of $T^{tt}$ over the 2D disk.
Violation relative to the initial net energy remains below 0.05\%.
In contrast to the angular momentum conservation, the energy conservation is well satisfied with only minor violation even without imposing it as a penalty term in the loss function.

\section{Numerical Demonstration of Mutual Spin-Orbit Conversion}\label{Sec:4}

We now provide the first numerical demonstration of the mutual conversion between spin and orbital angular momentum in the two limiting cases: (i) an initial state with purely orbital angular momentum and no spin polarization, which is analogous to the Barnett effect \cite{barnett1915magnetization} and is also relevant to noncentral heavy-ion collisions~\cite{Hattori:2019lfp,Fukushima:2020ucl,Gallegos:2021bzp,Li:2020eon, Hu:2021lnx,Hu:2022azy,Singh:2022ltu,Cao:2022aku,Daher:2022wzf,Sarwar:2022yzs,
Kiamari:2023fbe,Xie:2023gbo,Ren:2024pur,Florkowski:2017ruc,
Peng:2021ago,
Weickgenannt:2022zxs,
Weickgenannt:2023btk,Bhadury:2024ckc,
Montenegro:2017rbu,Montenegro:2020paq,
Hongo:2021ona, Hu:2021lnx,Hu:2022azy,Tiwari:2024trl,Florkowski:2024bfw,
Fang:2024skm,Fang:2024hxa,Fang:2024sym, Wagner:2024fry,
Dey:2024cwo, She:2024rnx,Huang:2024ffg, Drogosz:2024gzv,
Wagner:2024fhf,Chiarini:2024cuv,Daher:2025pfq,
Bhadury:2025fil,Sapna:2025yss,Singh:2025hnb,Abboud:2025shb}; and (ii) an initial state with purely spin polarization and no orbital angular momentum, which is analogous to the Einstein--de Haas effect~\cite{richardson1908mechanical,einstein1915experimental}.
All the quantities shown in this section are normalized by $\bar{e}$.

\subsection{Orbital to Spin Angular Momentum Conversion: Test Case for High-Energy Heavy-Ion Collisions}\label{Sec:4_o_to_s}

We discuss the first case where an initial orbital angular momentum is converted into spin polarization.
We adopt the initial conditions in Eqs.~\eqref{Eq:init_orb_1}--\eqref{Eq:init_orb_5} 
with the parameters summarized in Table~\ref{Tab:para_3_1}\@.
To isolate the role of the rotational viscosity $ \gamma$,
the key quantity in spin hydrodynamics,
we perform a control test with $ \gamma = 0$ for the ideal fluid
and compare the results with those for $ \gamma =2$.
Numerical results shown in this section are obtained after
50,000 iterations for $\gamma=0$ and 150,000 iterations for $\gamma=2$.
After these iterations, the loss function converges,
maintaining the total energy conserved with deviations from the initial value suppressed below 0.02\% for $\gamma=0$ and 0.05\% for $\gamma=2$.
The numerical setup for the $\gamma=2$ run is the same as that described in Sec.~\ref{Sec:3_Metho}, and the $\gamma=0$ run uses the same setup except for the value of $\gamma$.
The training takes approximately 1 hour for $\gamma = 0$ and 2 hours for $\gamma = 2$, using one NVIDIA H100 NVL GPU.

\begin{table}[htbp]
\begin{center}
\begin{tabular}{r|c}
\hline\hline
$\bar{e}$      & $1$ \\
$R$            & $1$ \\
$t_\text{max}$ & $0.4$ \\
$\gamma_\phi$  & $0,\ 2$ \\
$\tau_\phi$    & $\gamma_\phi$ \\
$\delta_1$     & $0.2$ \\
\hline\hline
\end{tabular}
\end{center}
\caption{
Parameters used in Sec.~\ref{Sec:4_o_to_s}, expressed in units of $\bar{e}$.
}
\label{Tab:para_3_1}
\end{table}

\begin{figure*}[tp]
\begin{minipage}{0.4\textwidth}
    \centering
    \includegraphics[width=\textwidth]{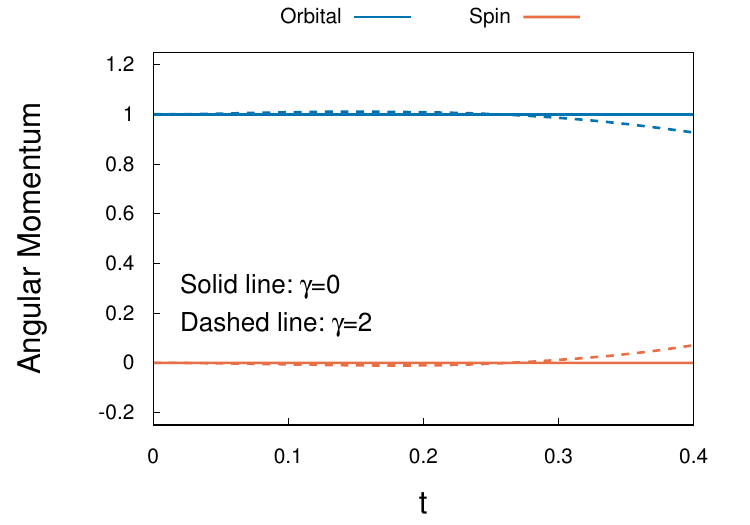}
\end{minipage}
\caption{
The time evolution of the
net orbital and spin angular momentum.
All quantities are normalized by the spatially integrated total angular momentum at the initial time $t=0$.
}
\label{Fig:ang_HIC}
\end{figure*}

\begin{figure*}[tp]
\begin{minipage}{0.33\textwidth}
    \centering
    \includegraphics[width=\textwidth]{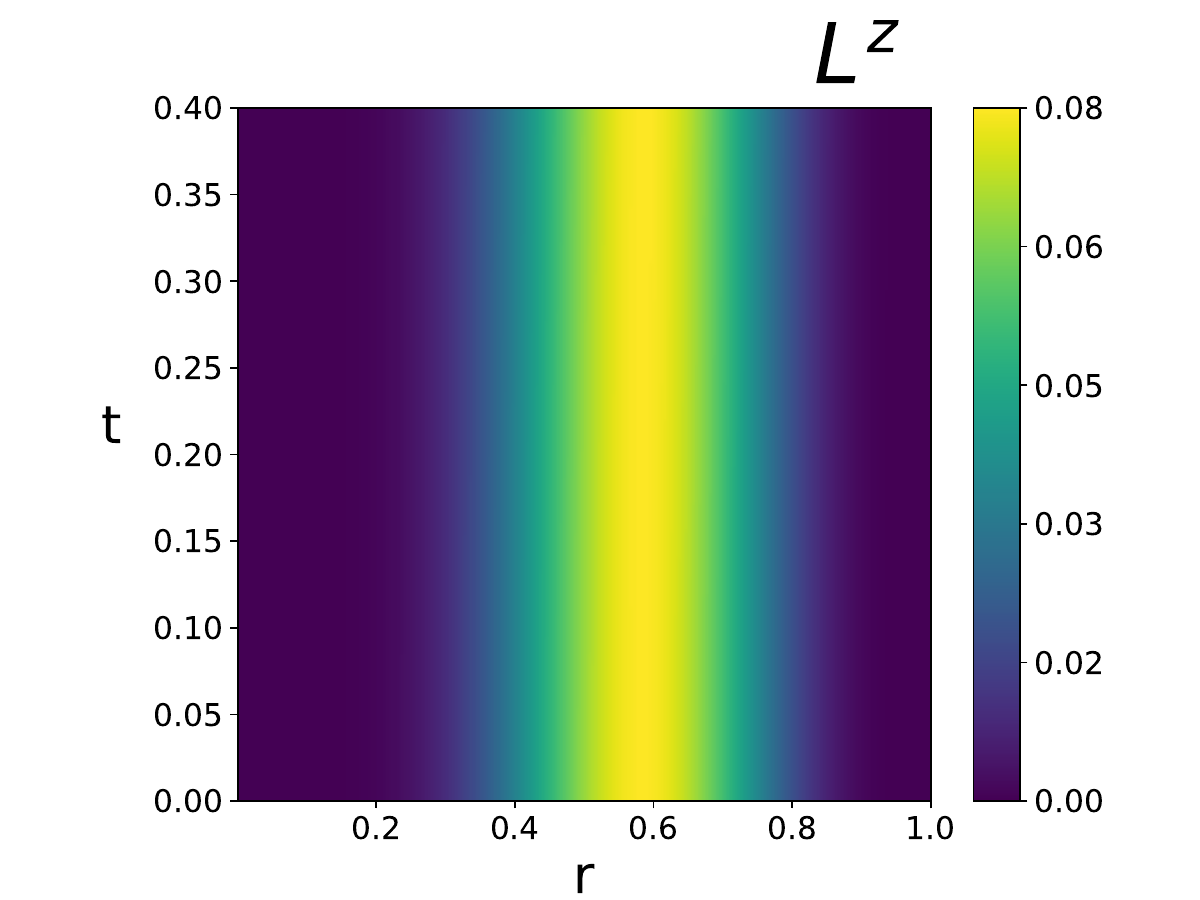}
\end{minipage}
\begin{minipage}{0.33\textwidth}
    \centering
    \includegraphics[width=\textwidth]{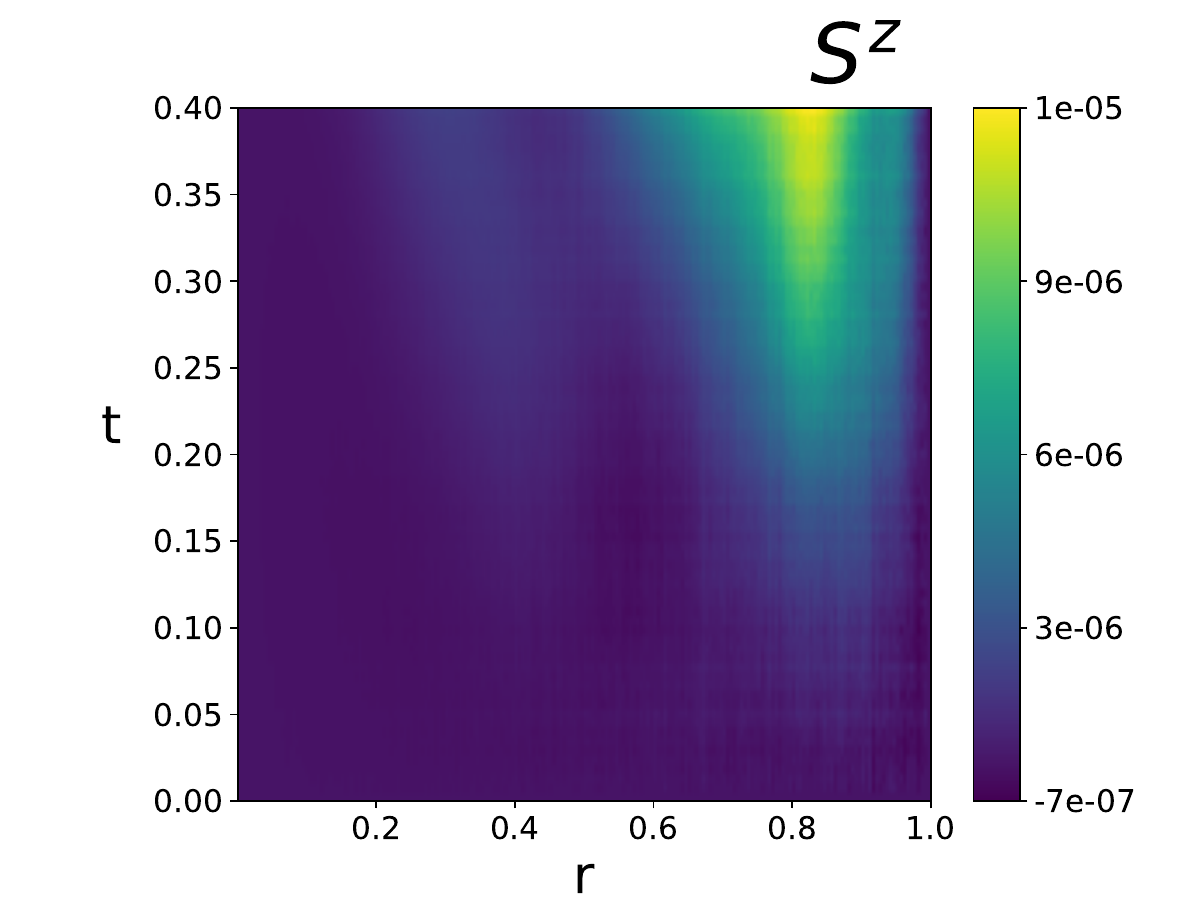}
\end{minipage}\\
\begin{minipage}{0.33\textwidth}
    \centering
    \includegraphics[width=\textwidth]{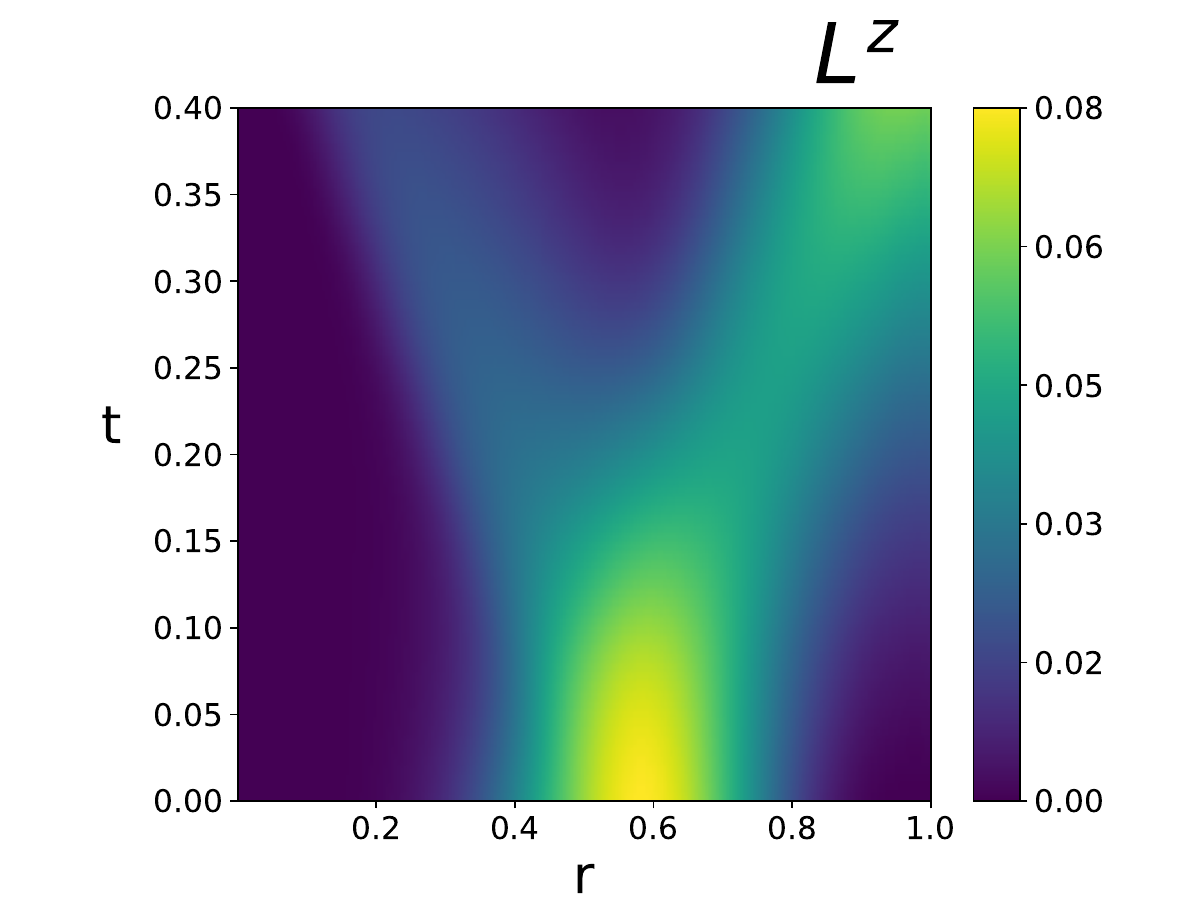}
\end{minipage}
\begin{minipage}{0.33\textwidth}
    \centering
    \includegraphics[width=\textwidth]{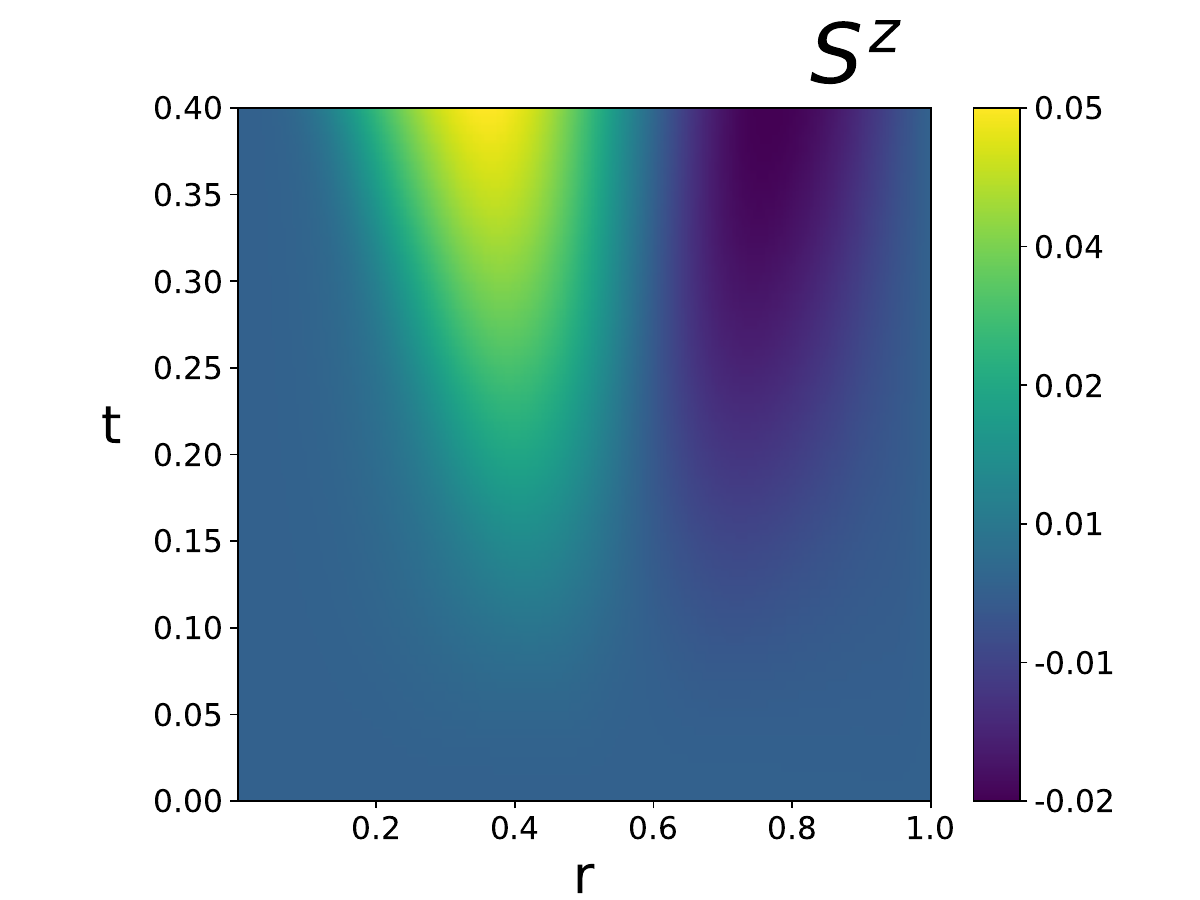}
\end{minipage}
\caption{
Heatmaps in the $t$-$r$ plane of the orbital and spin angular momentum density.
The upper and lower rows show those for $\gamma=0$ and $\gamma=2$, respectively.
}
\label{Fig:ang_heatmap_HIC}
\end{figure*}

\subsubsection{Orbital-to-spin conversion}

We show the time evolution of the orbital and spin angular momentum densities,
$L^z = x\,\Theta^{ty} - y\,\Theta^{tx} =r^2( 4/3 e u^t u^\varphi + u^r \phi^{r\varphi}/u^t )$ and $S^z$.
While Fig.~\ref{Fig:ang_HIC} shows the global values, i.e.,
the integral of these quantities over the spatial volume of the 2D disk,
Fig.~\ref{Fig:ang_heatmap_HIC} shows the local values without the integration.

In Fig.~\ref{Fig:ang_HIC}, we observe a sizable net conversion from orbital to spin angular momentum at $\gamma=2$, which is approximately 7\% of the initial orbital angular momentum within the shown time scale.
Importantly, global angular momentum conservation is maintained with high accuracy, with deviations controlled to less than 0.06\% and 0.4\% for $\gamma=0$ and $\gamma=2$, respectively.
Therefore, the conversion magnitude is approximately 20 times larger than the error stemming from the violation of the total angular momentum.
In the ideal-fluid case, where the conversion is turned off ($\gamma=0$), the
conversion rate is less than 0.02\% of the total angular momentum.
This deviation is of the same order as the 0.06\% error in the total angular momentum and is considered numerical artifacts.

\begin{figure*}[tp]
    \centering
    \includegraphics[width=0.5\textwidth]{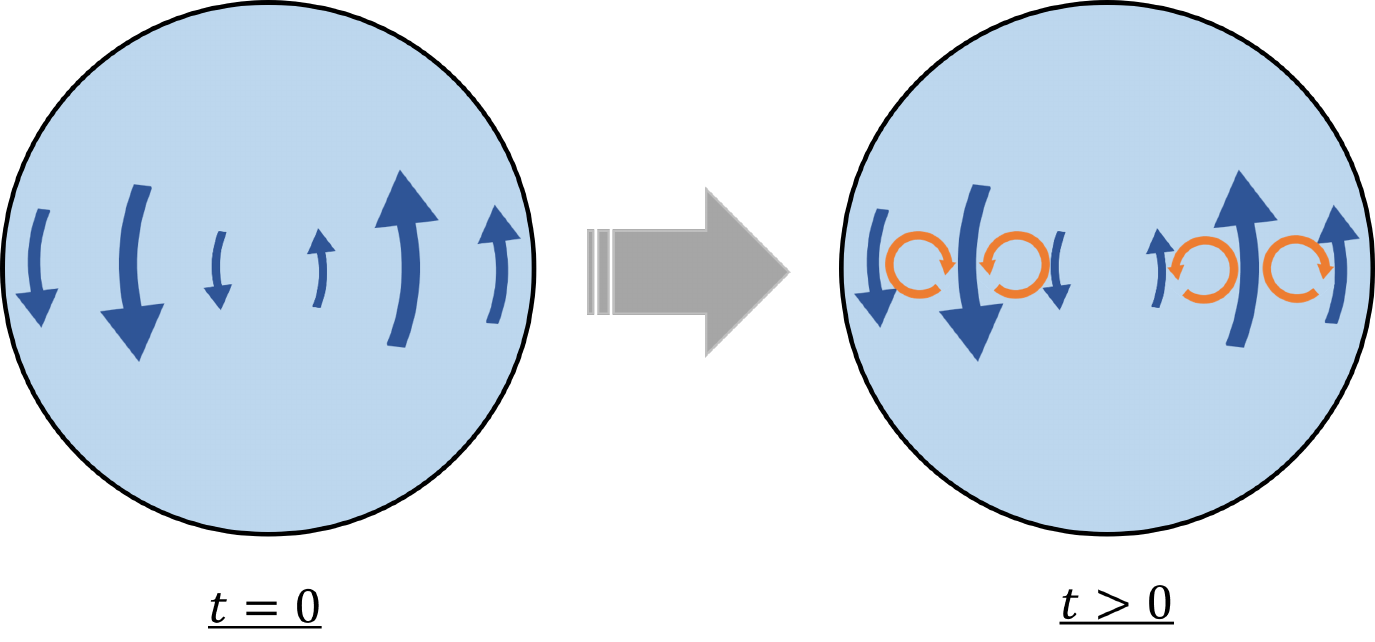}
\caption{Conceptual sketch: spin angular momentum (orange arrows) induced by initial orbital angular momentum (blue arrows).}
 \label{Fig:orb_to_spin}
\end{figure*}

\begin{figure*}[tp]
\begin{minipage}{0.33\textwidth}
    \centering
    \includegraphics[width=\textwidth]{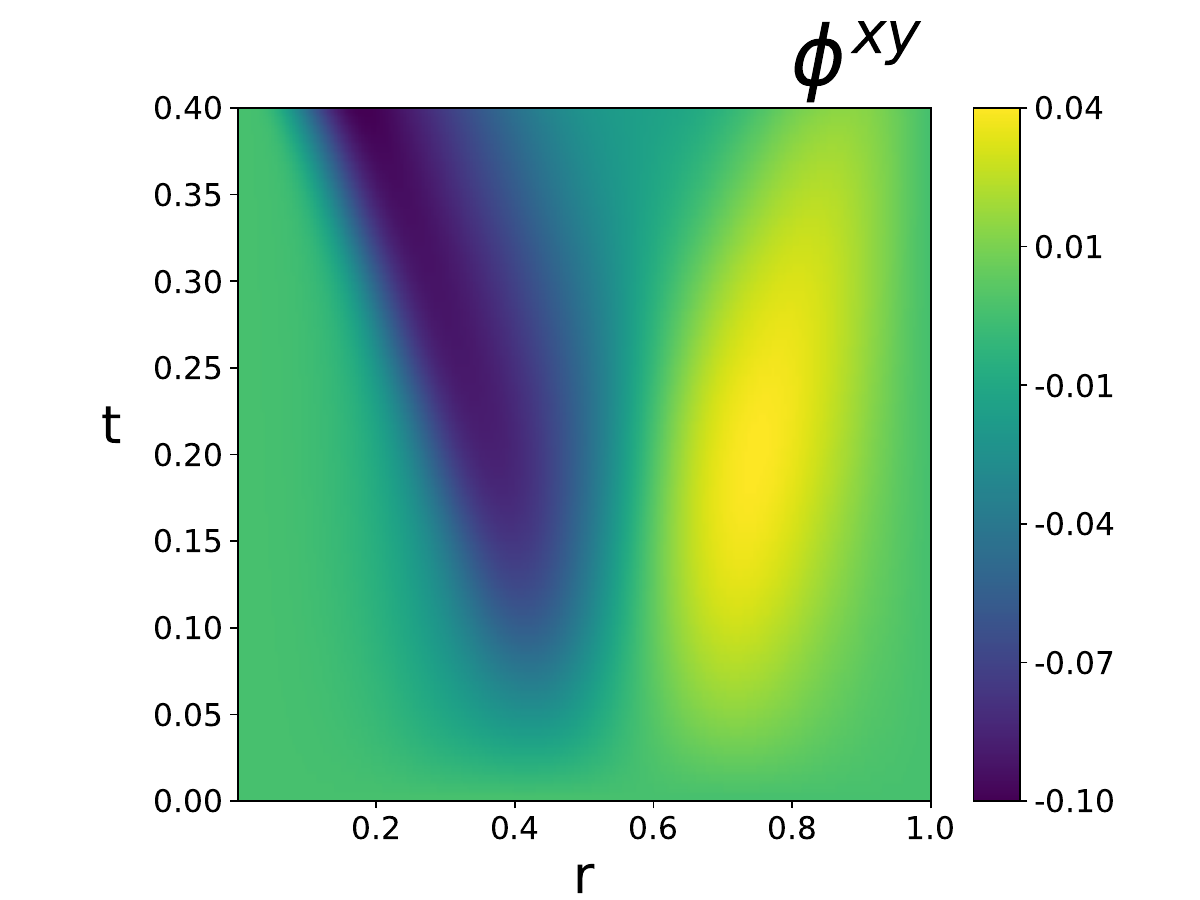}
\end{minipage}
\begin{minipage}{0.33\textwidth}
    \centering
    \includegraphics[width=\textwidth]{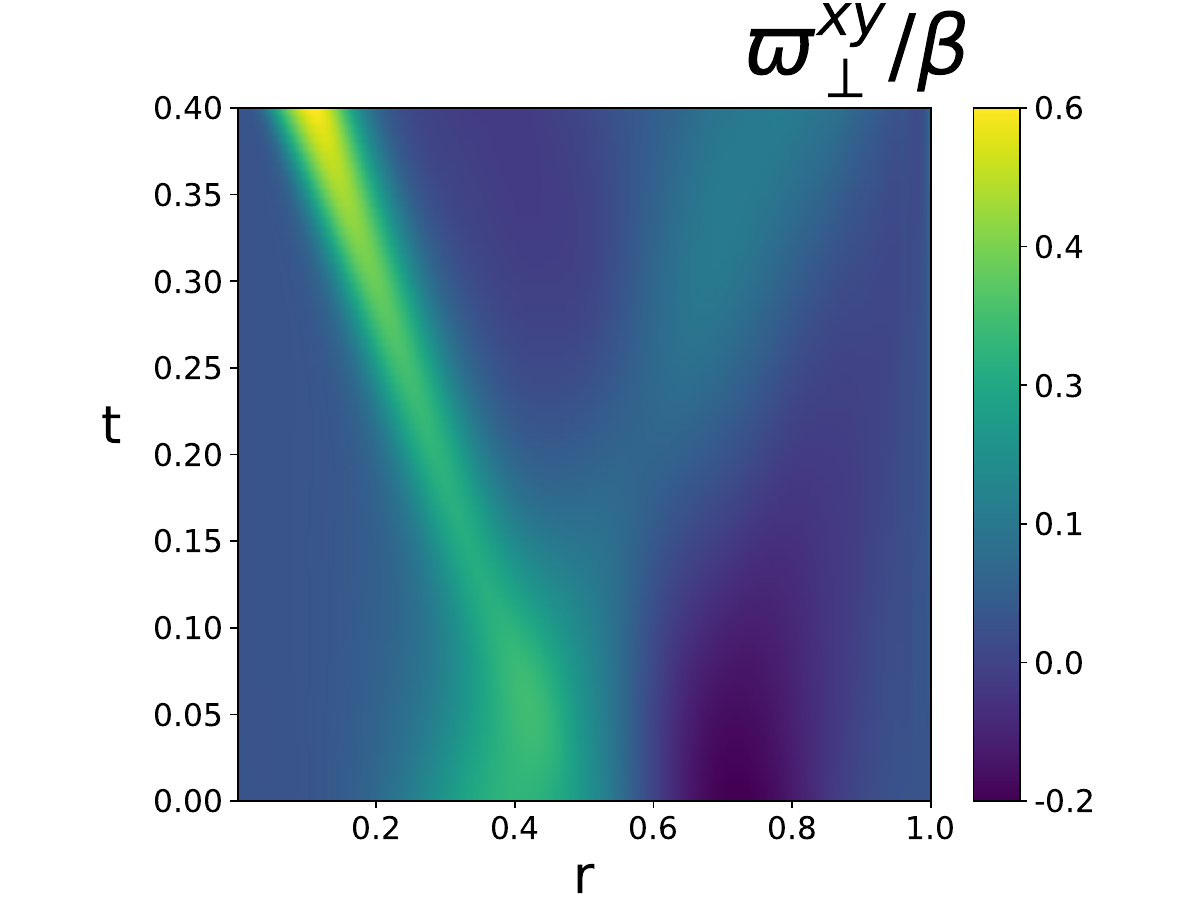}
\end{minipage}
\begin{minipage}{0.33\textwidth}
    \centering
    \includegraphics[width=\textwidth]{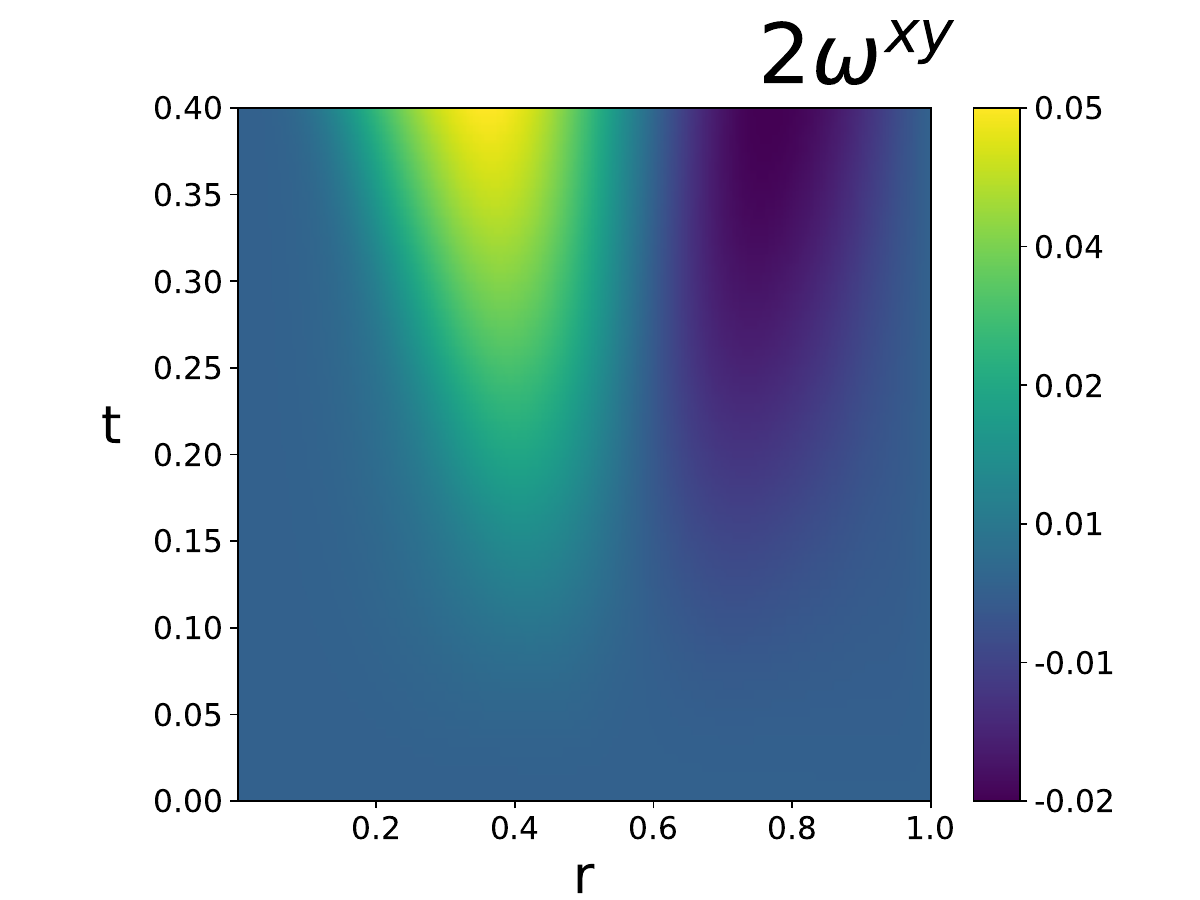}
\end{minipage}
\caption{
Heatmaps in the $t$-$r$ plane of
the $xy$ components of the couple-stress tensor, the transverse thermal vorticity, and the spin potential,
shown from left to right, for the case of $\gamma=2$.
}
\label{Fig:phi_HIC}
\end{figure*}

Figure~\ref{Fig:ang_heatmap_HIC} presents heatmaps of the local orbital and spin angular momentum densities in the
$t$-$r$ plane.
In the upper panels for the ideal fluid ($\gamma=0$), both orbital and spin
components remain essentially unchanged, with variations of order $10^{-5}$--$10^{-4}$,
which is smaller than the error in the total angular momentum conservation.
In the lower panels, a finite rotational viscosity ($ \gamma=2$) acts to redistribute the local orbital and spin angular momentum densities, as well as the global quantities found in Fig.~\ref{Fig:ang_HIC}.
In the time evolution of orbital angular momentum, the initial peak structure near $ r=0.5$ splits into two, with the lower and higher peaks migrating inward and outward in the cylinder, respectively.
A corresponding structure emerges in the spin component shown in the lower right panel.
Namely, positive and negative spin polarization emerges in the interior and exterior regions, respectively.

The time evolution of the local quantities in Fig.~\ref{Fig:ang_heatmap_HIC} can be understood
based on a development of the rotation-rate mismatch $\rho^{\mu\nu}$.
The spin-orbit conversion rate is proportional to $ \rho^{\mu\nu}$
that appears in the first-order spin evolution equation (\ref{Eq:const-spin-orbit}) and is given by
the difference between
the transverse thermal vorticity $ \varpi^{\mu \nu}_\perp $ and the spin potential $ 2\beta \omega^{\mu \nu}$ as shown in Eq.~(\ref{eq:rho}).
The local density profiles found in Fig.~\ref{Fig:ang_heatmap_HIC}
are basically generated by the mismatch between these two measures of orbital motion and spin.
This draws an intuitive picture in Fig.~\ref{Fig:orb_to_spin};
the initial rotational flow (blue arrows) activates the rotational viscous effect with position-dependent signs at the initial time $ \rho^{\mu \nu} (t=0)=\beta^{-1} \varpi^{\mu \nu}_\perp $ that in turn creates spin density (orange arrows).

In the MIS extension for causal hydrodynamics, it is more legitimate to
refer to the couple-stress tensor $ \phi^{\mu\nu}$, which relaxes to $ \rho^{\mu\nu}$
following the relaxation equation~\eqref{Eq:hydro5}.
We numerically examine the dynamics of $ \phi^{\mu\nu}$ to support the aforementioned picture in Fig.~\ref{Fig:orb_to_spin}.
In Eq.~\eqref{Eq:hydro5}, the early-time evolution of $\phi^{\mu \nu}$ is governed by $\rho^{\mu \nu}$ as $ \phi^{\mu \nu}(t=0)=0$ in the initial condition (\ref{Eq:init_orb_5}).
In Fig.~\ref{Fig:phi_HIC}, we show the $xy$ components of the couple-stress tensor $\phi^{\mu\nu}$, the transverse thermal vorticity $ \varpi^{xy}_\perp $, and the spin potential $\omega^{xy}$ for $\gamma=2$.
Since these three tensors are totally antisymmetric, their $xy$ components are represented as the $r\varphi$ components multiplied by $r$.

According to the flow initial condition given in Eq.~\eqref{Eq:init_orb},
the transverse thermal vorticity $ \varpi^{xy}_\perp $
in the early-time evolution has positive and negative values
in the interior and exterior of the cylinder, respectively,
while $\omega^{xy}$ remains negligibly small.
Therefore, the early-time dynamics of $\phi^{xy}$
is strongly correlated with the transverse thermal vorticity $ \varpi^{xy}_\perp $ as observed in the left two panels.
At later times, the rotational viscous effect creates
the spin density $S^z$ as seen in Eq.~\eqref{Eq:hydro4}
with a two-peak structure similar to that of $\phi^{xy}$ but with the opposite signs.
That is, the transverse thermal vorticity $ \varpi^{xy}_\perp $
and the spin potential $\omega^{xy}$ develop similar structures to resolve their mismatch and diminish the rotational viscous effect accordingly.

\begin{figure*}[tp]
\begin{minipage}{0.33\textwidth}
    \centering
    \includegraphics[width=\textwidth]{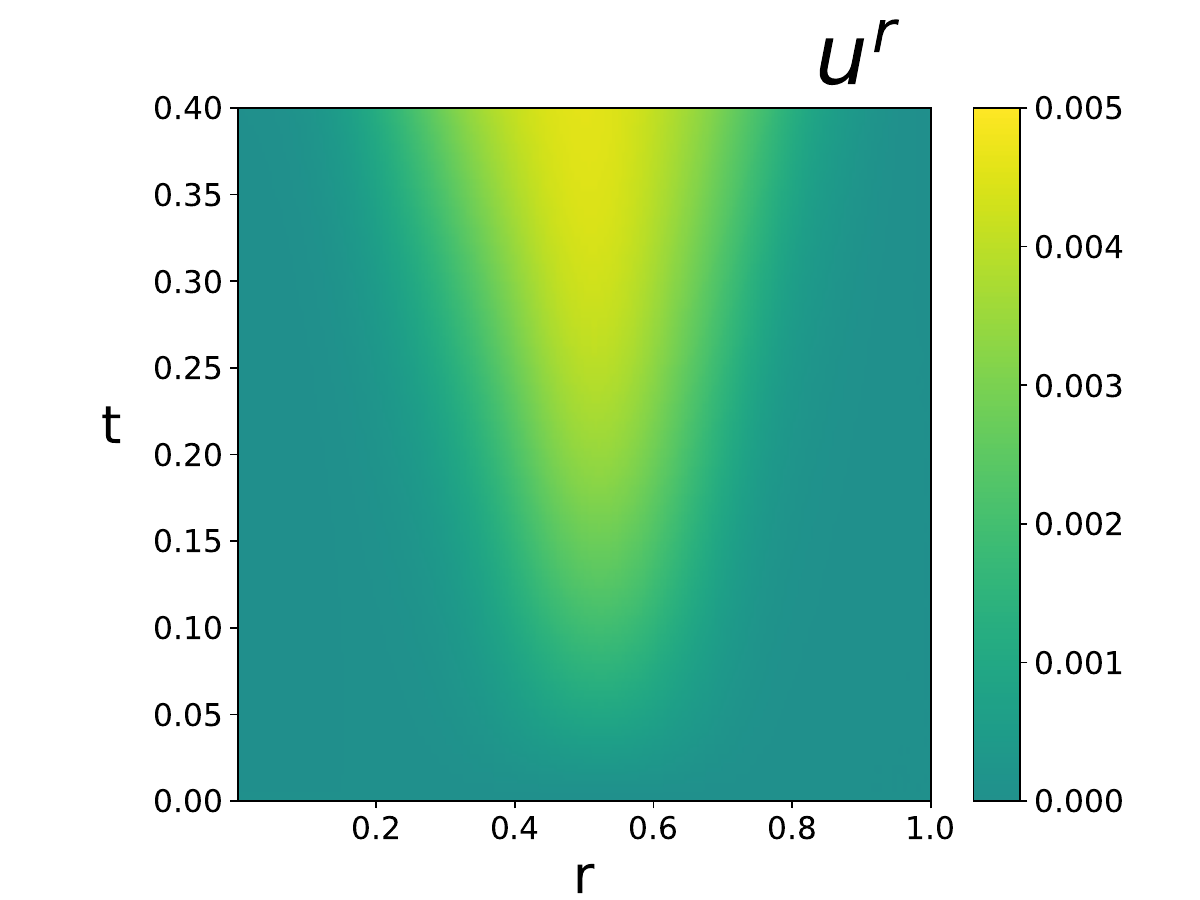}
\end{minipage}
\begin{minipage}{0.33\textwidth}
    \centering
    \includegraphics[width=\textwidth]{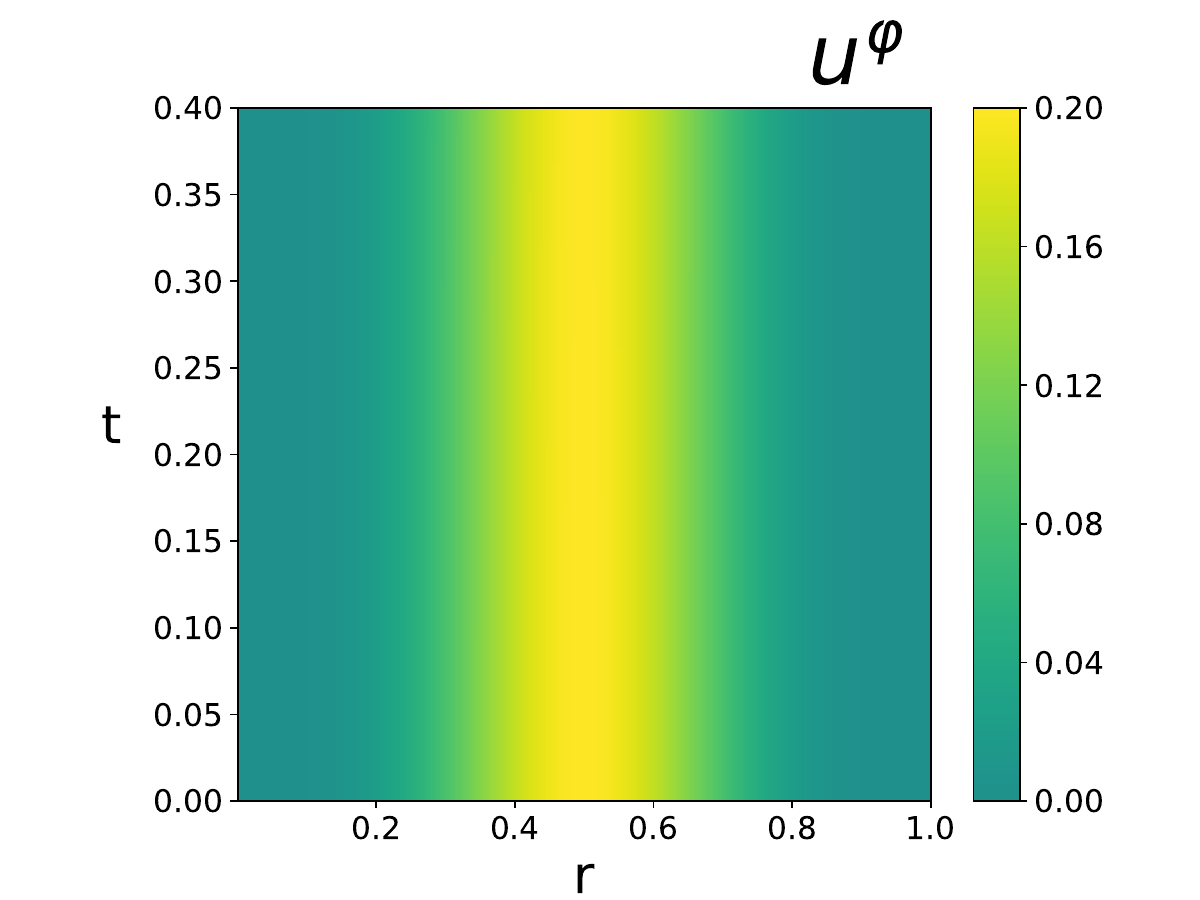}
\end{minipage}
\begin{minipage}{0.33\textwidth}
    \centering
    \includegraphics[width=\textwidth]{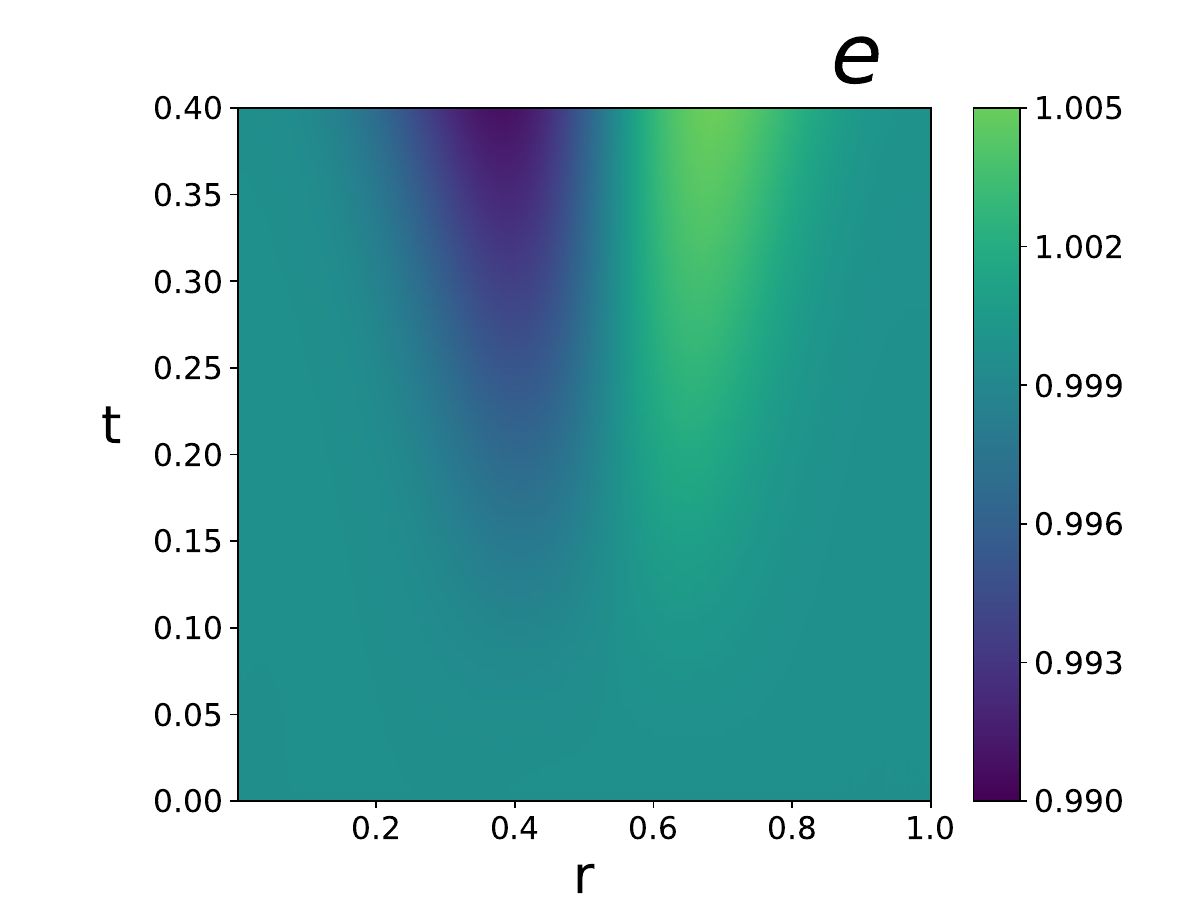}
\end{minipage}
\begin{minipage}{0.33\textwidth}
    \centering
    \includegraphics[width=\textwidth]{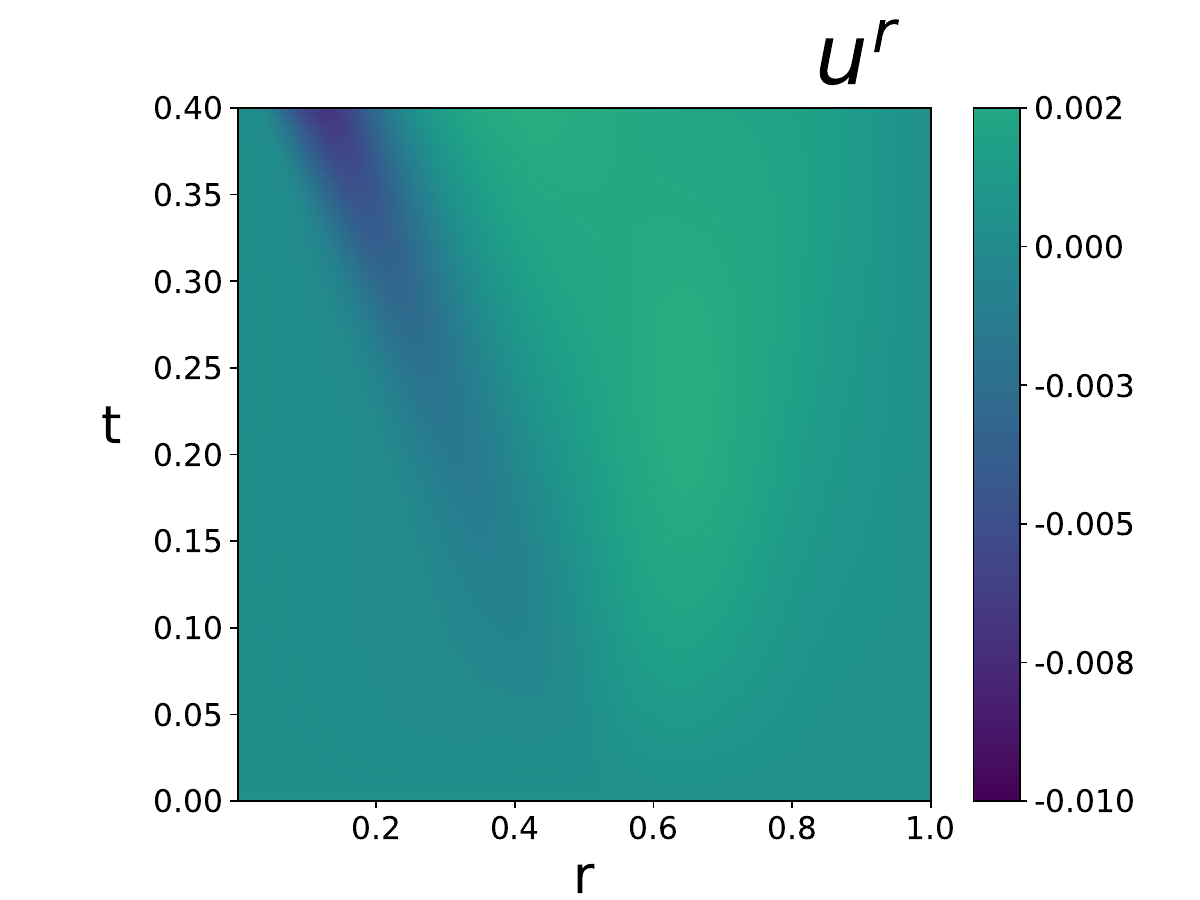}
\end{minipage}
\begin{minipage}{0.33\textwidth}
    \centering
    \includegraphics[width=\textwidth]{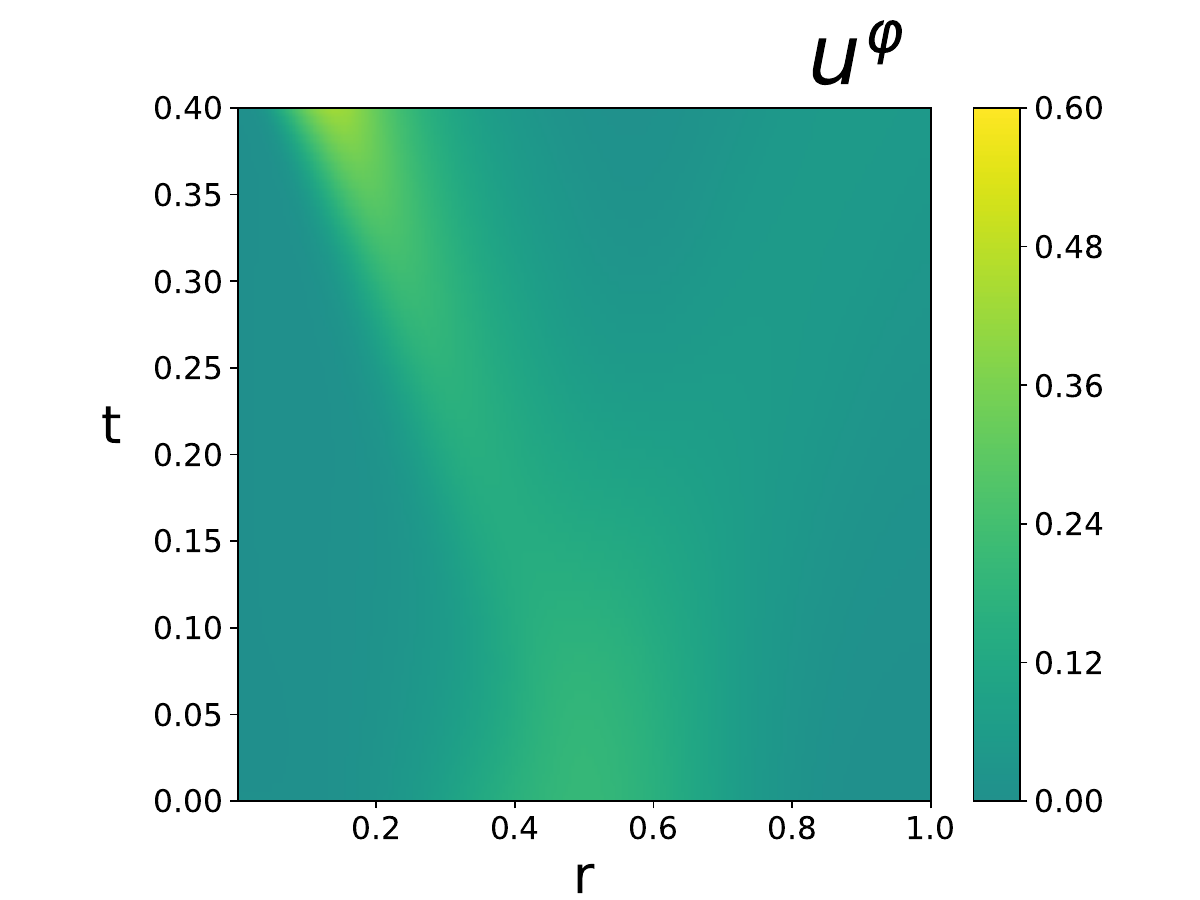}
\end{minipage}
\begin{minipage}{0.33\textwidth}
    \centering
    \includegraphics[width=\textwidth]{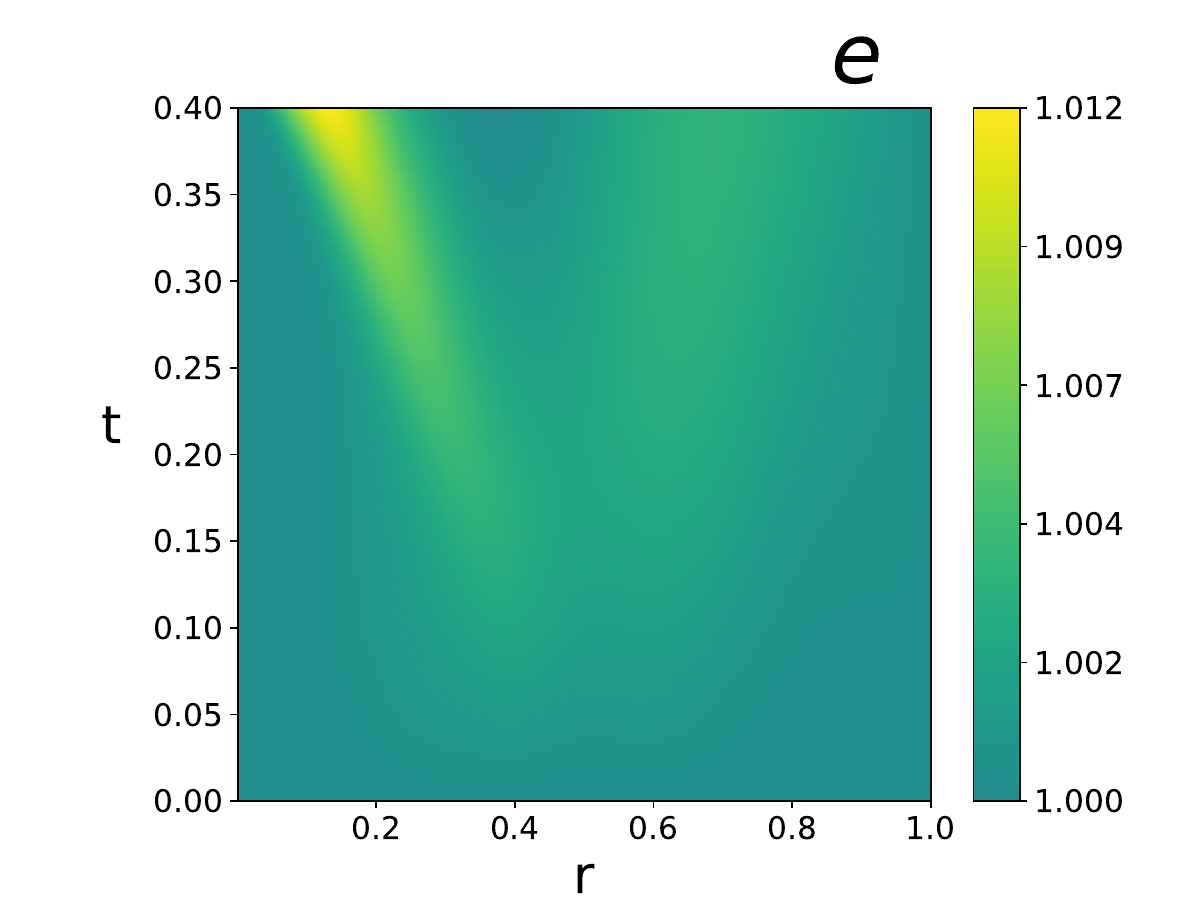}
\end{minipage}
\caption{
Heatmaps in the $t$-$r$ plane of the radial velocity $u^r$, the angular velocity $u^\varphi$,
and the thermodynamic energy density $e$, shown from left to right,
for $\gamma=0$ in the upper row and $\gamma=2$ in the lower row.
}
\label{Fig:ene_HIC}
\end{figure*}

\subsubsection{Feedback on the flow profile}

The initial rotational flow induces the finite spin angular momentum in the
fluid through the Barnett effect, but it further alters the initial rotational
flow and the energy distribution.  We here demonstrate such feedback to the
time evolution of the flow and energy density by comparing the simulation
results for $\gamma = 2$ and the ideal fluid ($\gamma = 0$).

In Fig.~\ref{Fig:ene_HIC}, we show the heatmaps in the $t$-$r$ plane of the radial velocity $u^r$, the angular velocity $u^\varphi$,
and the thermodynamic energy density $e$.

The radial velocity $u^r$ (left panels) receives strong influence from the dynamics of the angular momentum.
In the ideal fluid case ($\gamma=0$), the flow is oriented outward everywhere in the cylinder, consistent with centrifugal expansion.
In contrast, for $\gamma=2$, inward motion also develops, suggesting that rotational viscosity drives a significant redistribution of radial flow.
As we further elaborate below, the angular velocity and energy density also exhibit significant modifications by the rotational viscous effect.

In the case of an ideal fluid ($\gamma=0$), one can understand the flow evolution with the effect of the centrifugal force.
The radial flow $u^r$ is absent at the initial time, but gradually develops a peak structure near $r\sim 0.5$ as the system evolves.
In contrast, the angular velocity $u^{\varphi}$ remains nearly unchanged from its initial profile.
Near $r \simeq 0.5$, where both $r u^{\varphi}$ and $u^{r}$ reach their maximum values,
the absolute value of $ru^{\varphi}$ is approximately twenty times
larger than
$u^{r}$, showing that the dynamics is dominated by the rotational component rather than the radial one.
The thermodynamic energy density, initially uniform in $r$, decreases for $r < 0.5$ and increases for $r > 0.5$ as the system evolves.
This is consistent with
the effect of the centrifugal force.

The formation of the peak structure in $u^r$ is understood by the centrifugal force.
In fact, the ideal hydrodynamic equation for $u^r$, $D u^r \sim r (u^\varphi)^2 + \cdots$,
contains a term analogous to the centrifugal force in the Newton equation $dv^r/dt = r (\omega^\varphi)^2$ for the radial velocity $v^r$ and angular velocity $\omega^\varphi$ of a rotating rigid body.
This term is expected to play an important role in the development of the peak structure since the angular velocity $u^\varphi$ is much larger than the radial velocity $u^r$ in the present setup.

The evolution in the spinful fluid with $\gamma=2$ exhibits
qualitatively different features from those of the ideal fluid,
arising from the competition between the ideal fluid behavior and rotational viscous effects.
In particular, the radial profile of $u^\varphi$ is significantly modified from its initial distribution,
whereas in the ideal fluid it remains almost unchanged.
Initially, $u^{\varphi}$ has a single peak structure at $r = 0.5$,
but in the viscous case this peak splits into two,
with one migrating toward smaller $r$ and the other toward larger $r$.
Furthermore, the energy density develops a positive peak in the deep interior of the cylinder,
which overrides the naive expectation based on the centrifugal force alone.
These features indicate that the rotational viscous effect not only mediates orbital-to-spin conversion but also leaves a clear imprint on the collective flow and thermodynamic structure of the spinful fluid.

\subsection{Spin to Orbital Angular Momentum Conversion: The Einstein--de Haas Effect}\label{Sec:4_s_to_o}

\begin{figure*}[tp]
\begin{minipage}{0.4\textwidth}
    \centering
    \includegraphics[width=\textwidth]{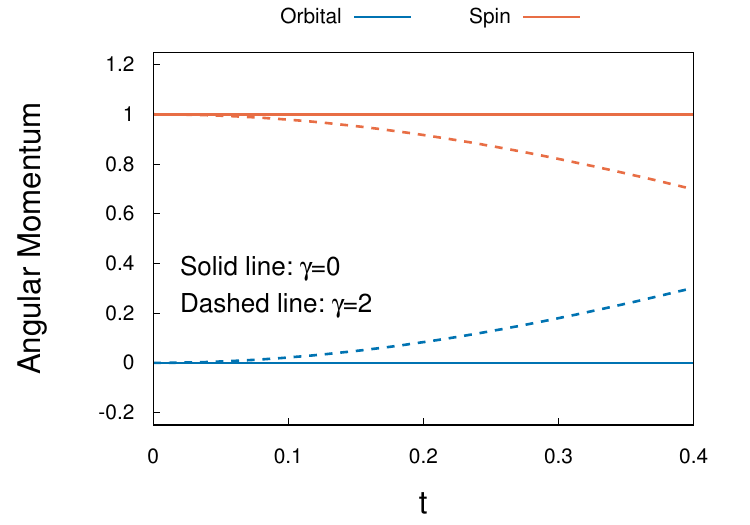}
\end{minipage}
\caption{
Time evolution of the global orbital and spin angular momentum
for $\gamma = 2$, shown with a trivial case at $ \gamma=0$.
All quantities are normalized by the spatially integrated total angular momentum at the initial time $t=0$.
}
\label{Fig:ang_EdH}
\end{figure*}

\begin{figure*}[tp]
\begin{minipage}{0.33\textwidth}
    \centering
    \includegraphics[width=\textwidth]{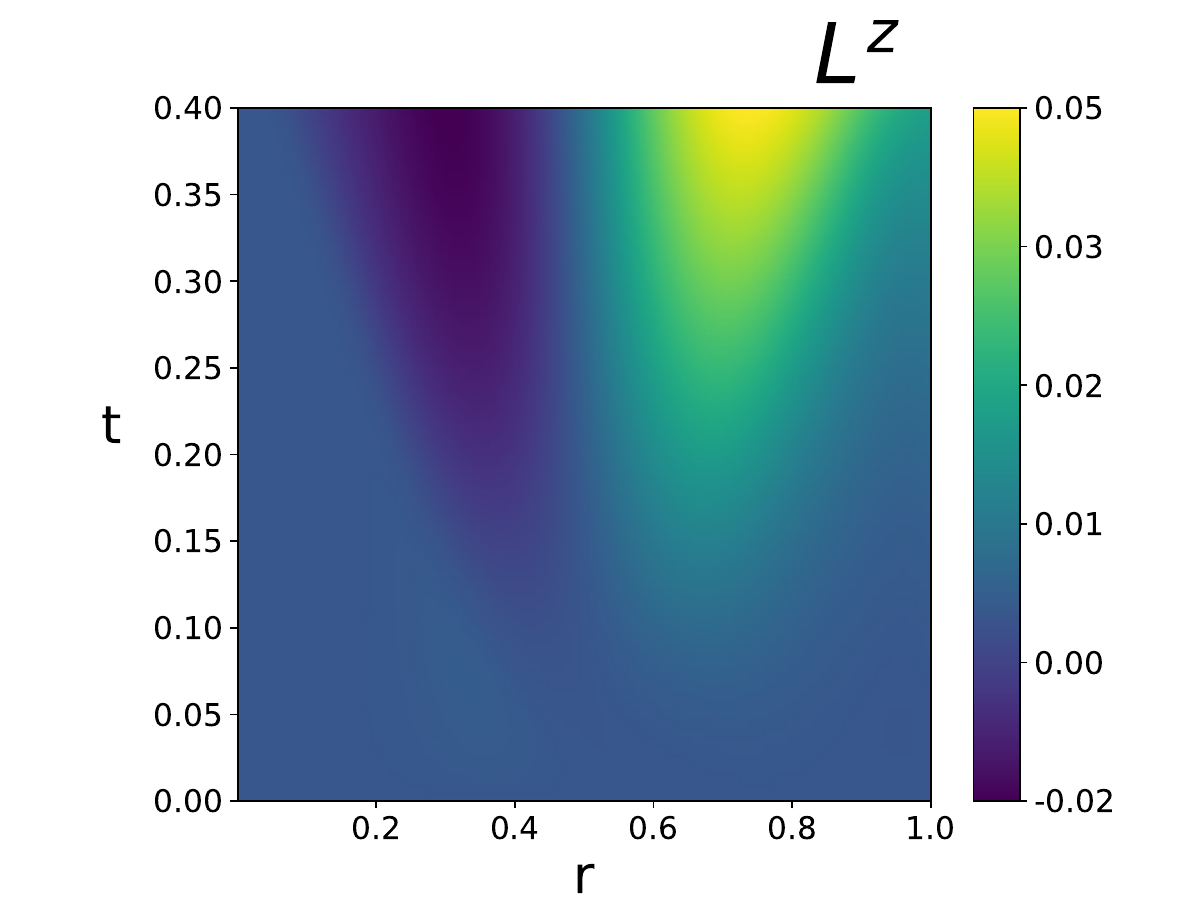}
\end{minipage}
\begin{minipage}{0.33\textwidth}
    \centering
    \includegraphics[width=\textwidth]{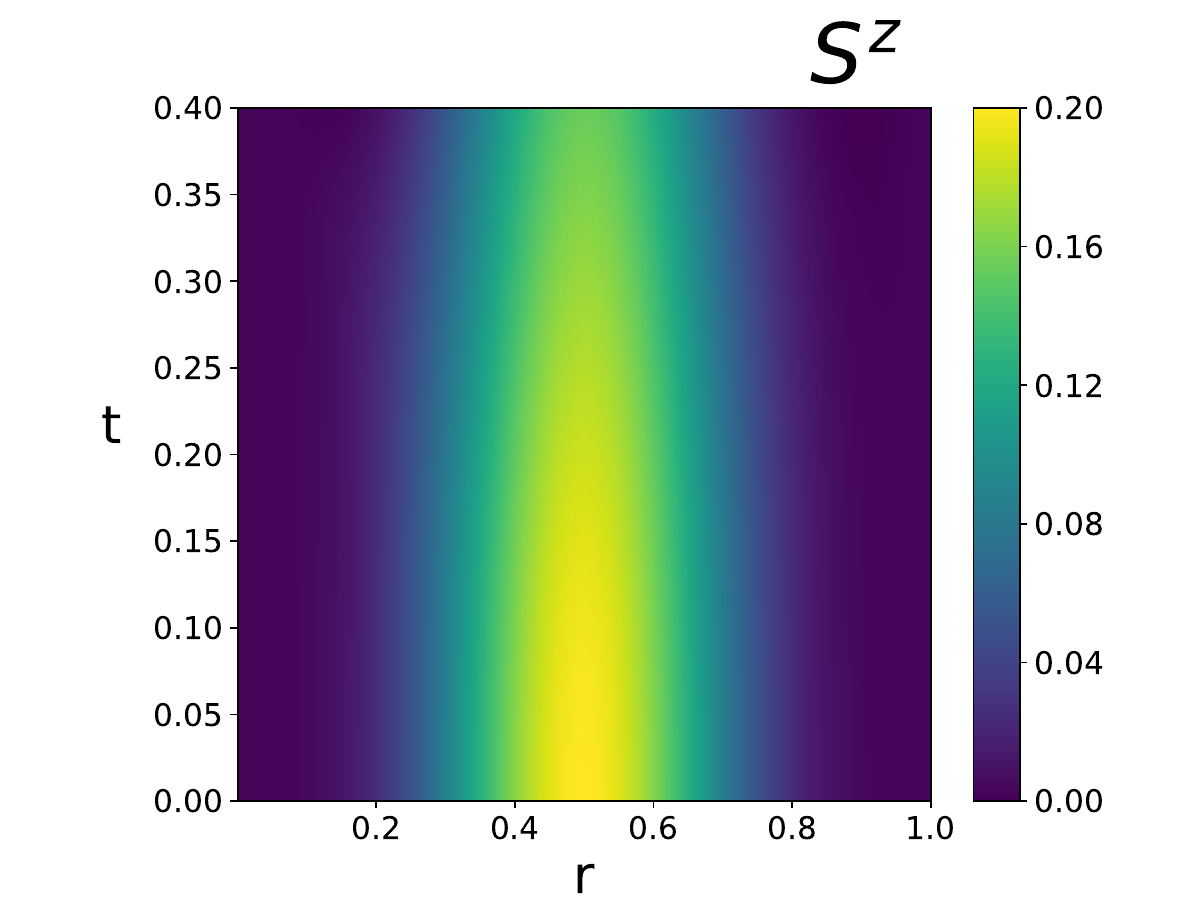}
\end{minipage}
\caption{
Heatmaps in the $t$-$r$ plane of orbital and spin angular momentum, shown from left to right, for $\gamma=2$.
}
\label{Fig:ang_heatmap_EdH}
\end{figure*}

We investigate the conversion of spin polarization to orbital angular momentum, in the opposite way to that discussed in the previous subsection.
This is the second case mentioned in the beginning of this section and is an analog of the Einstein-de Haas effect~\cite{richardson1908mechanical,einstein1915experimental},
in which spin polarization in a magnetized material
is converted into mechanical rigid rotation of the material through microscopic interactions.
In the context of spin hydrodynamics,
rotational fluid motion is driven by the initial spin polarization in the presence of the rotational viscous effect.

We perform the simulation with a nonzero spin polarization
and no orbital angular momentum at the initial time.
The initial conditions are provided as
\begin{subequations}
\begin{align}
e(t=0,r) &= \bar{e}\ ,\\
u^r(t=0,r) &= 0\ ,\\
u^{\varphi}(t=0,r) &= 0\ ,\\
S^z(t=0,r) &= \delta_2 \cdot \sin^4 \left( \frac{\pi r}{R} \right)\ ,\\
\phi^{r\varphi}(t=0,r) &= 0\ ,
\end{align}
\end{subequations}
where $\delta_2$ is a parameter controlling the magnitude of the initial spin density.
The parameters used in the simulation are summarized in Table~\ref{Tab:para_3_2}\@.
Unlike in the previous subsection, we do not perform the simulations for the ideal fluid ($\gamma=0$)
because under the present initial conditions, which correspond to a global equilibrium configuration, all the hydrodynamic variables obviously remain unchanged in the ideal fluid dynamics.
Numerical results shown in this subsection are obtained after 150,000 iterations.
After these iterations, the loss function converges,
maintaining the total energy conserved with deviations from the initial value suppressed below 0.01\%. 
The numerical setup is the same as that described in Sec.~\ref{Sec:3_Metho}\@. The training
takes approximately 2 hours,
using one NVIDIA H100 NVL GPU.

\begin{table}[htbp]
\begin{center}
\begin{tabular}{r|c}
\hline\hline
$\bar{e}$      & $1$ \\
$R$            & $1$ \\
$t_\text{max}$ & $0.4$ \\
$\gamma_\phi$  & $2$ \\
$\tau_\phi$    & $2$ \\
$\delta_2$     & $0.2$ \\
\hline\hline
\end{tabular}
\end{center}
\caption{
Parameters used in Sec.~\ref{Sec:4_s_to_o}, expressed in units of $\bar{e}$.
}
\label{Tab:para_3_2}
\end{table}

\begin{figure*}[tp]
    \centering
    \includegraphics[width=0.5\textwidth]{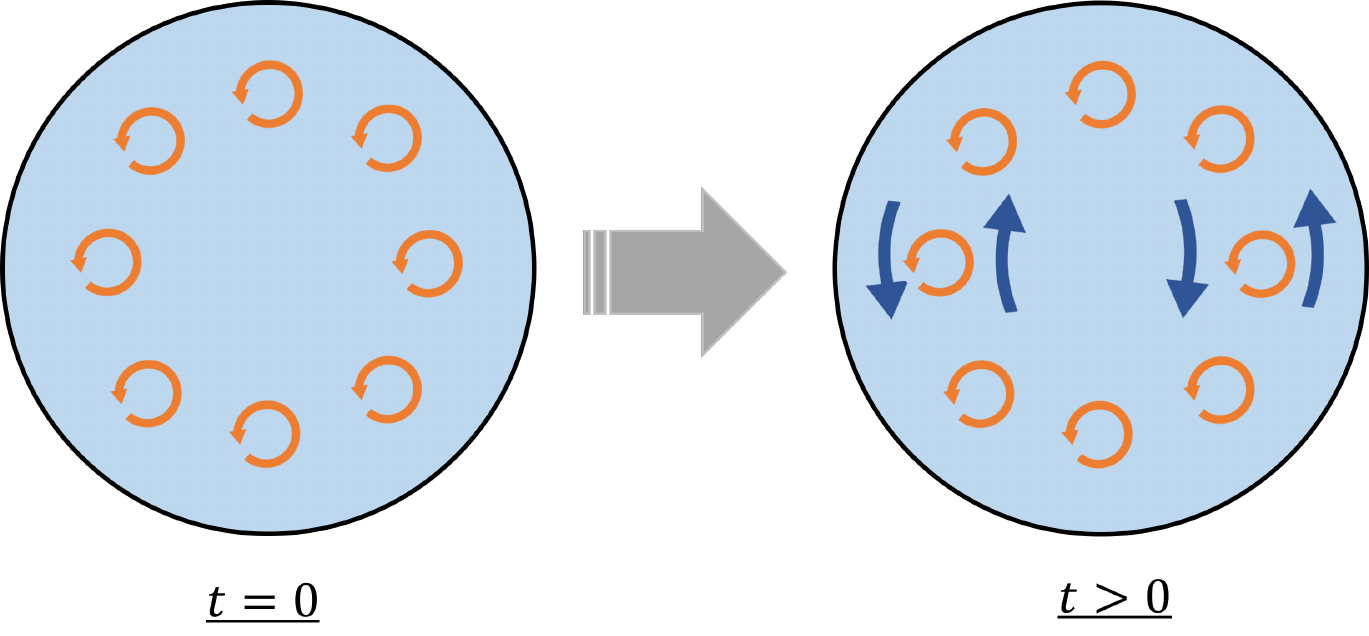}
\caption{Conceptual sketch: orbital angular momentum induced by initial spin angular momentum}
 \label{Fig:spin_to_orb}
\end{figure*}

\begin{figure*}[tp]
\begin{minipage}{0.33\textwidth}
    \centering
    \includegraphics[width=\textwidth]{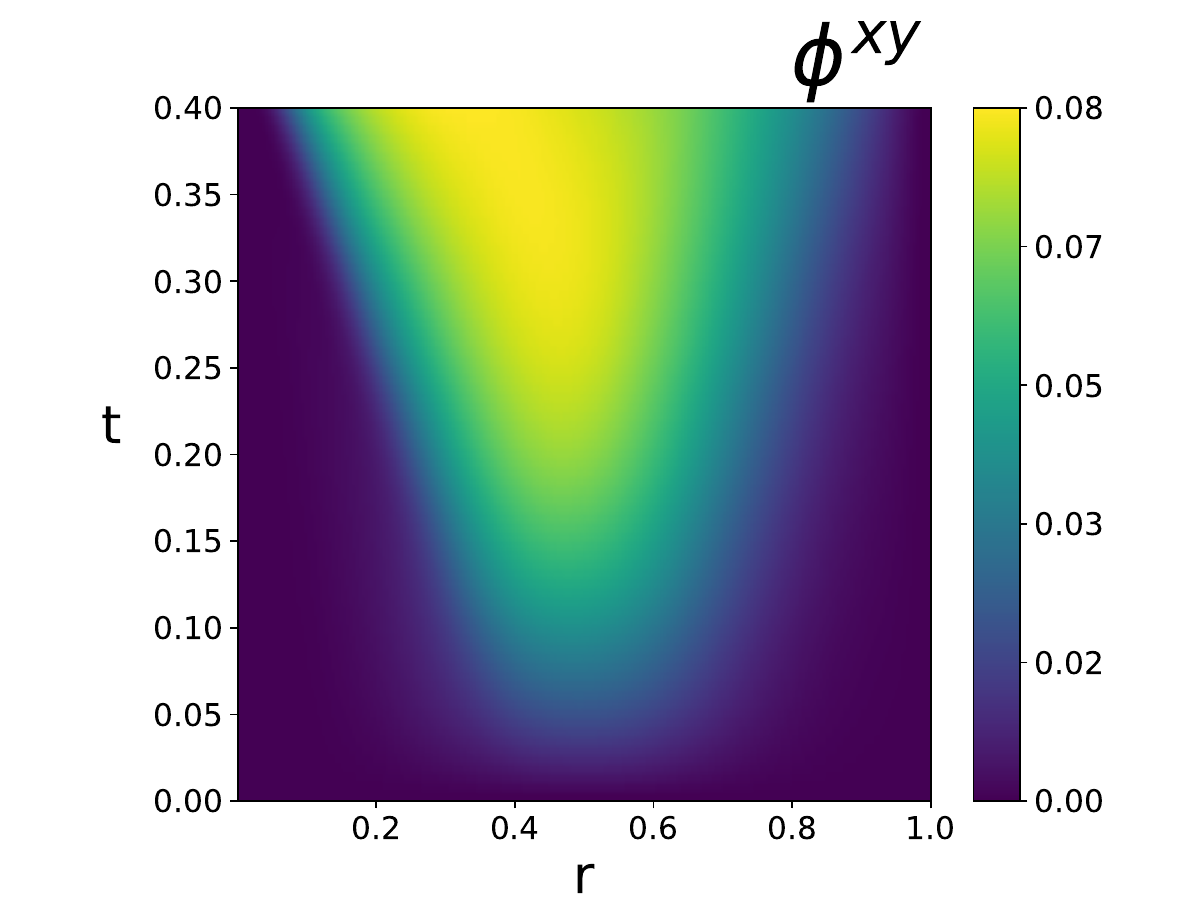}
\end{minipage}
\begin{minipage}{0.33\textwidth}
    \centering
    \includegraphics[width=\textwidth]{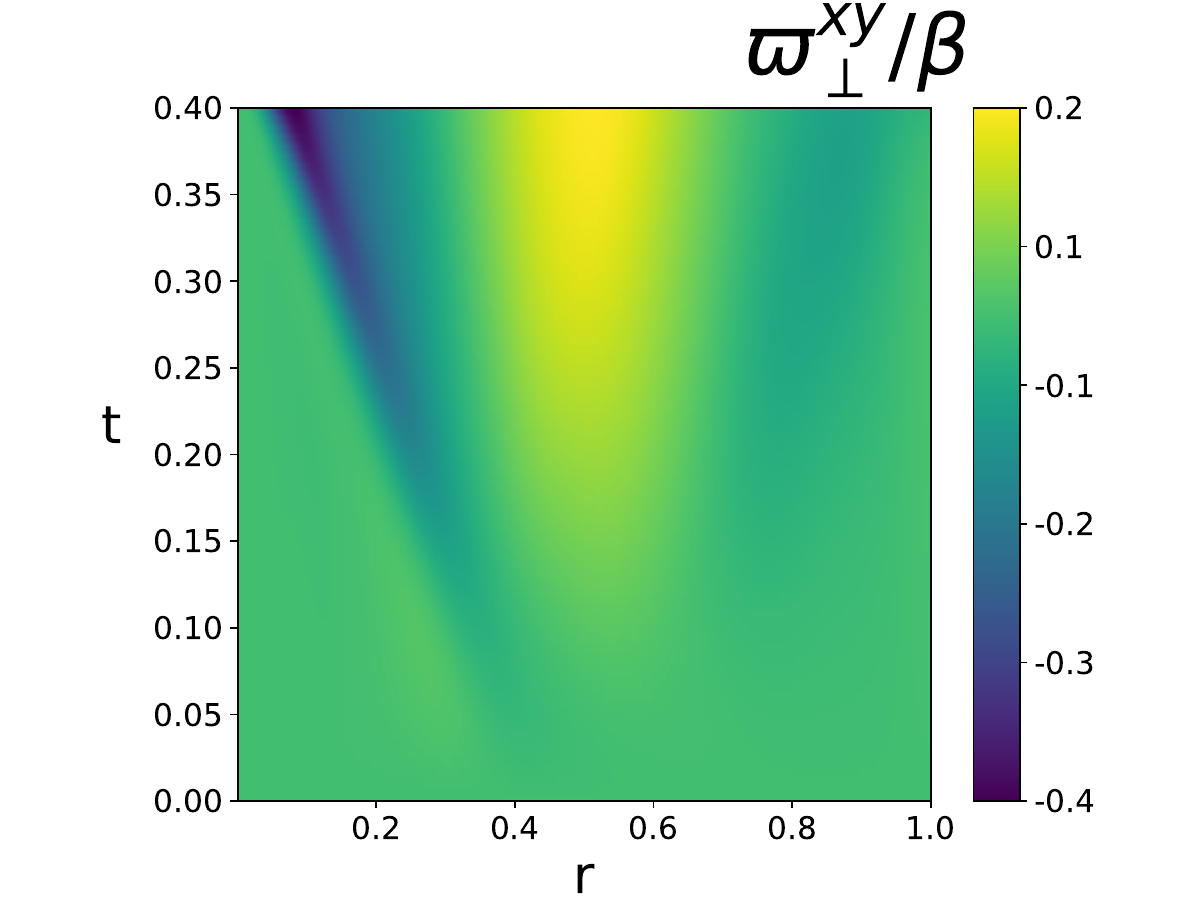}
\end{minipage}
\begin{minipage}{0.33\textwidth}
    \centering
    \includegraphics[width=\textwidth]{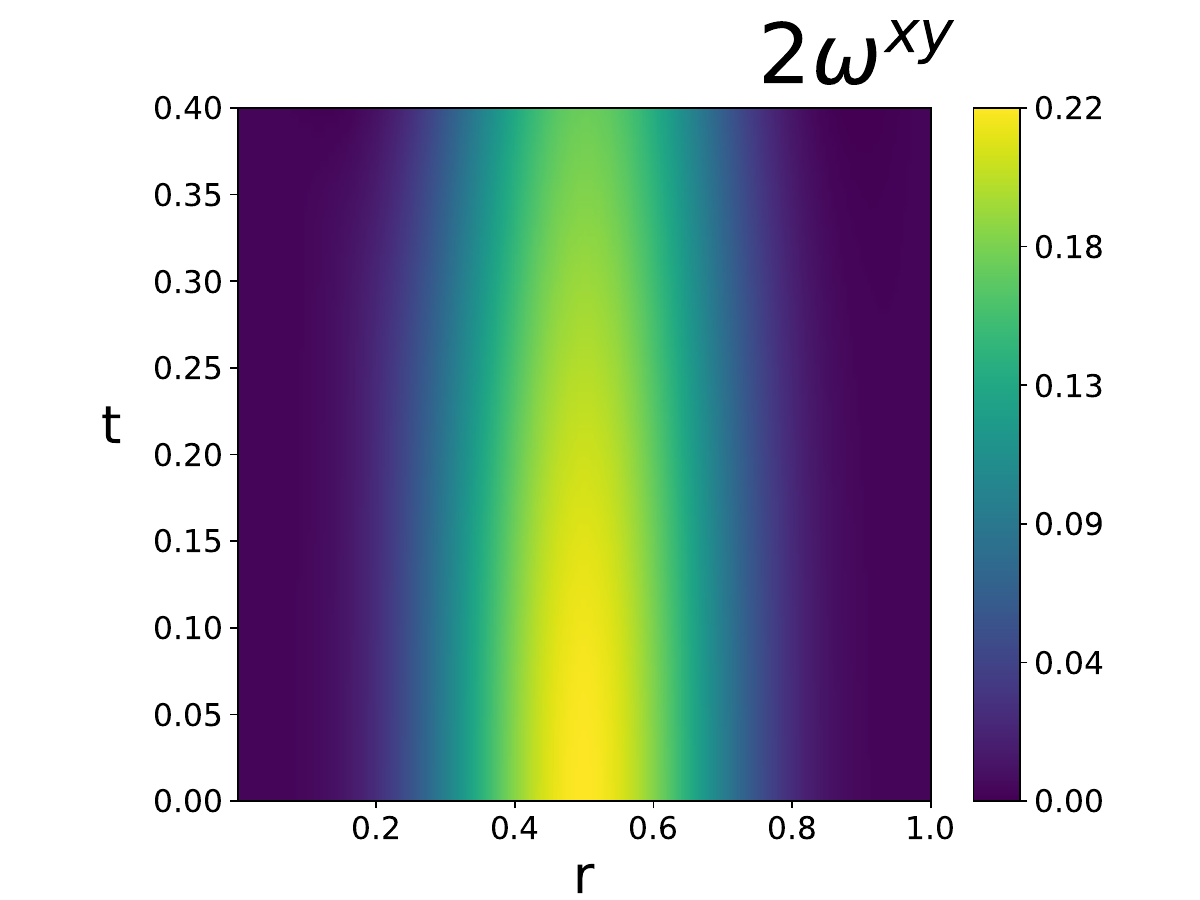}
\end{minipage}
\caption{
Heatmaps in the $t$-$r$ plane of
the $xy$ components of the couple-stress tensor, the transverse thermal vorticity, and the spin potential,
shown from left to right, for $\gamma=2$.
}
\label{Fig:phi_EdH}
\end{figure*}

\subsubsection{Spin-to-orbital conversion}

As in the previous subsection, we first show the numerical results
for the global and local angular momentum, and the rotation-rate mismatch $\rho^{\mu\nu}$ just below, and then the impact of the rotational viscous effect on the flow profile in Sec.~\ref{sec:Feedback-flow-EdH}.

In Fig.~\ref{Fig:ang_EdH}, we show the time evolution of the
global orbital and spin angular momentum, obtained by spatial integration of $L^z$ and $S^z$ over the 2D disk.
Our numerical simulations indicate spin-to-orbital conversion for approximately 30\% of the initial spin, with global angular momentum accurately conserved within 0.05\%.
The conversion magnitude is approximately 600 times larger than the error stemming from the violation of the total angular momentum conservation.
The ideal fluid case with $\gamma=0$ is a trivial static solution for a reference.
The comparison between these results clearly demonstrates the spin-to-orbital conversion by the rotational viscous effect.

In Fig.~\ref{Fig:ang_heatmap_EdH}, we show the heatmaps in the $t$-$r$ plane of orbital and spin angular momentum for $\gamma=2$.
The spin density $S^z$, which initially has a large positive magnitude, gradually disappears as the system evolves.
This behavior agrees with the decrease of the global value of spin angular momentum in Fig.~\ref{Fig:ang_EdH}.
On the other hand, the orbital angular momentum $L^z$ develops negative and positive values in the interior and exterior regions of the cylinder, respectively.
The magnitude of the positive component is slightly larger than that of the negative one,
so that the global value of orbital angular momentum increases as shown in Fig.~\ref{Fig:ang_EdH}.

As we analyzed the orbital-to-spin conversion in the previous subsection, the spin-to-orbital conversion is also understood within the framework of first-order spin hydrodynamics, in terms of the rotational viscous effect induced by the mismatch between the transverse thermal vorticity $ \varpi^{\mu \nu}_\perp $ and the spin potential $\omega^{\mu \nu}$.
This picture is illustrated in Fig.~\ref{Fig:spin_to_orb}, where the initial spin of the internal degrees of freedom (orange arrows) induces a rotational viscous effect with position-dependent sign of $\rho^{\mu \nu} = -2\omega^{\mu \nu}$ at the initial time $t=0$. In turn, the rotational viscous effect generates a position-dependent rotational flow (blue arrows) at later times.

To perform a more legitimate analysis, we evaluate
the couple-stress tensor $ \phi^{\mu\nu}$ as in the previous subsection.
In Fig.~\ref{Fig:phi_EdH}, we show the $xy$ components of the couple-stress tensor $\phi^{xy}$,
the transverse thermal vorticity $\varpi^{xy}_\perp$, and the spin potential $\omega^{xy}$ for $\gamma=2$.
The couple-stress tensor $\phi^{xy}$ is zero in the initial state,
but starts to develop a peak near $r=0.5$ after $t \sim 0.02$.
This spatial profile is correlated with that of the spin potential $\omega^{xy}$ concentrated near $r=0.5$;
since $\omega^{xy}$ is much larger than $\varpi^{xy}_\perp$ in the early-time dynamics at $t<0.02$,
this mismatch gives rise to the initial growth of $\phi^{xy}$.
Thus, both $\omega^{xy}$ and $\phi^{xy}$ have peak structures centered near $r=0.5$.
Shortly afterward, the transverse thermal vorticity $\varpi^{xy}_\perp$ starts to develop negative and positive peaks in response to the rotational viscous correction $-2\gamma\rho^{\mu\nu}$.
The development of these opposite rotational flows is consistent with the picture in Fig.~\ref{Fig:spin_to_orb}.
Such a phenomenon never occurs in an ideal fluid, and is characteristic to spin hydrodynamics driven by the rotational viscosity $\gamma$.

\begin{figure*}[tp]
\begin{minipage}{0.33\textwidth}
    \centering
    \includegraphics[width=\textwidth]{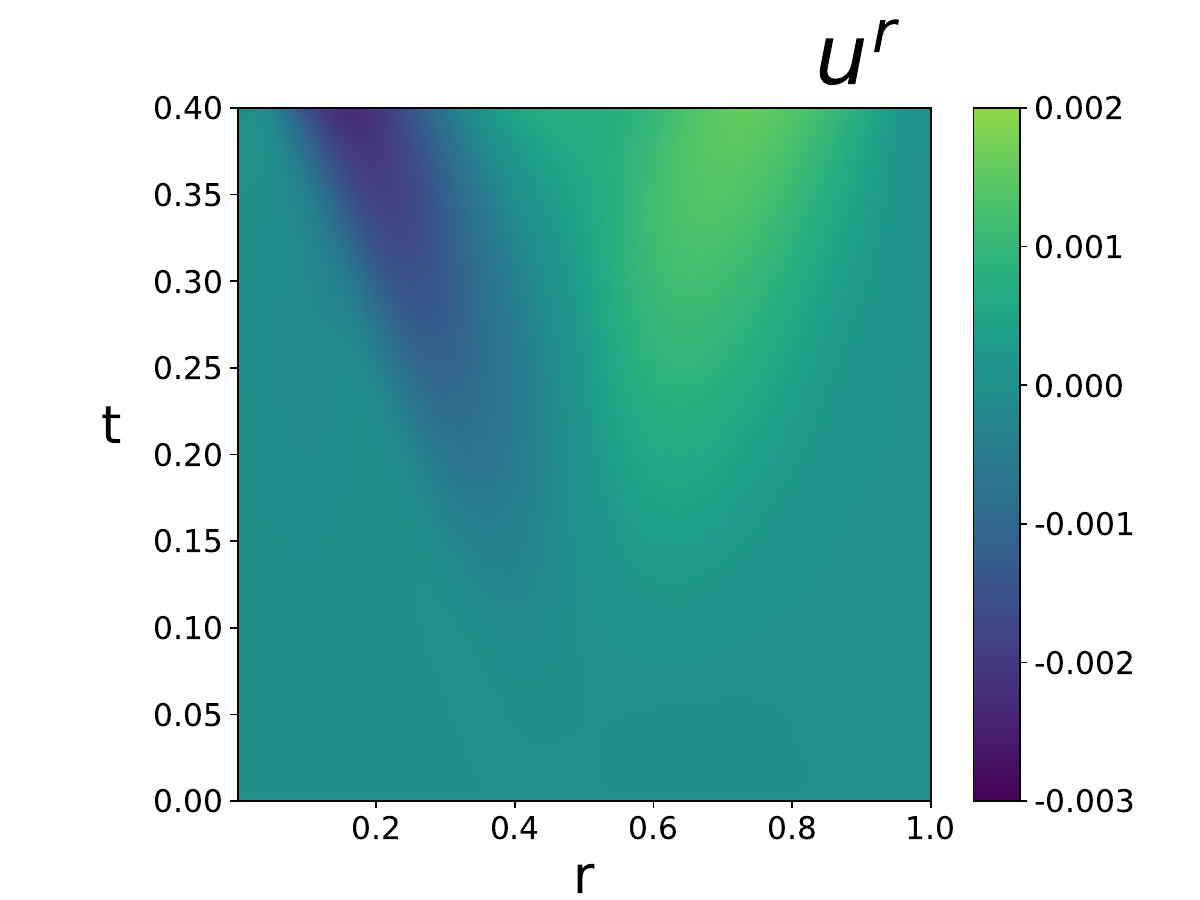}
\end{minipage}
\begin{minipage}{0.33\textwidth}
    \centering
    \includegraphics[width=\textwidth]{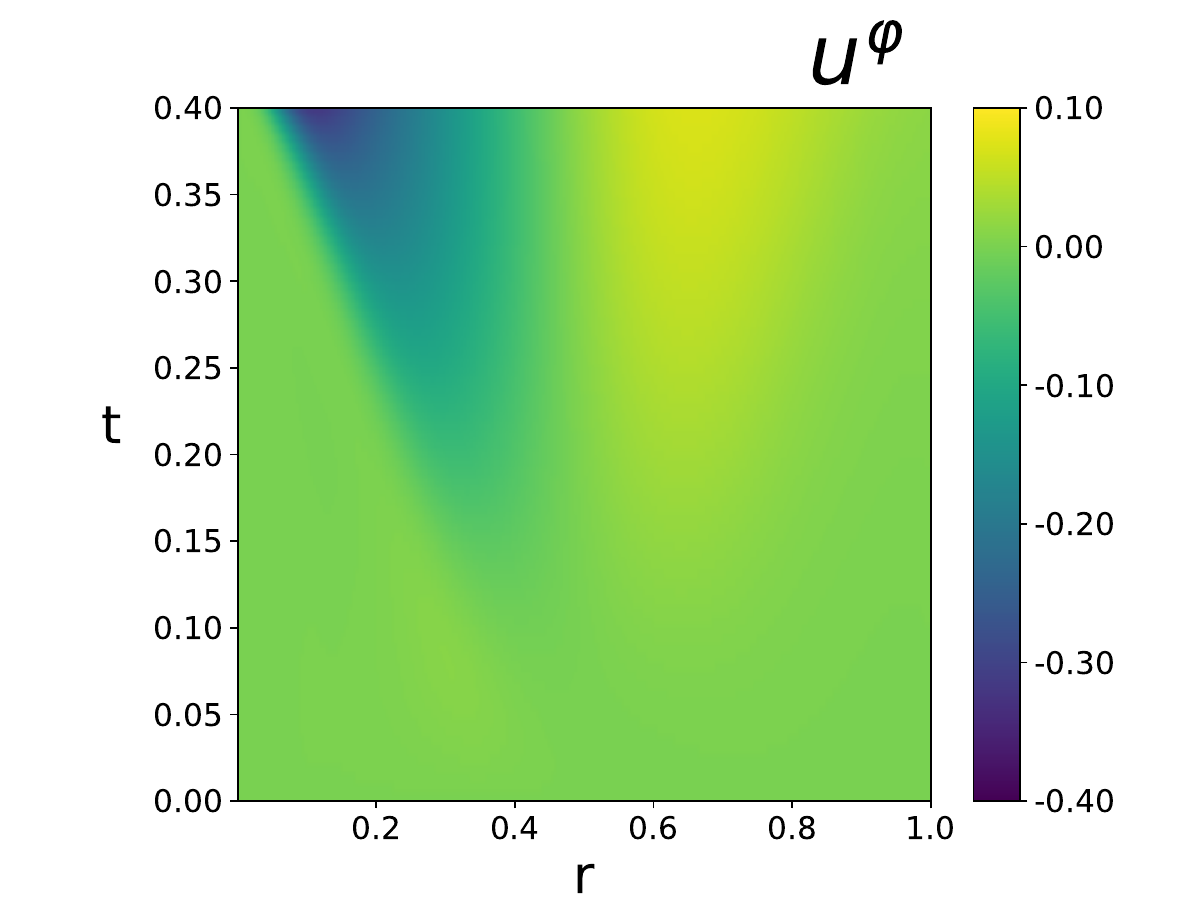}
\end{minipage}
\begin{minipage}{0.33\textwidth}
    \centering
    \includegraphics[width=\textwidth]{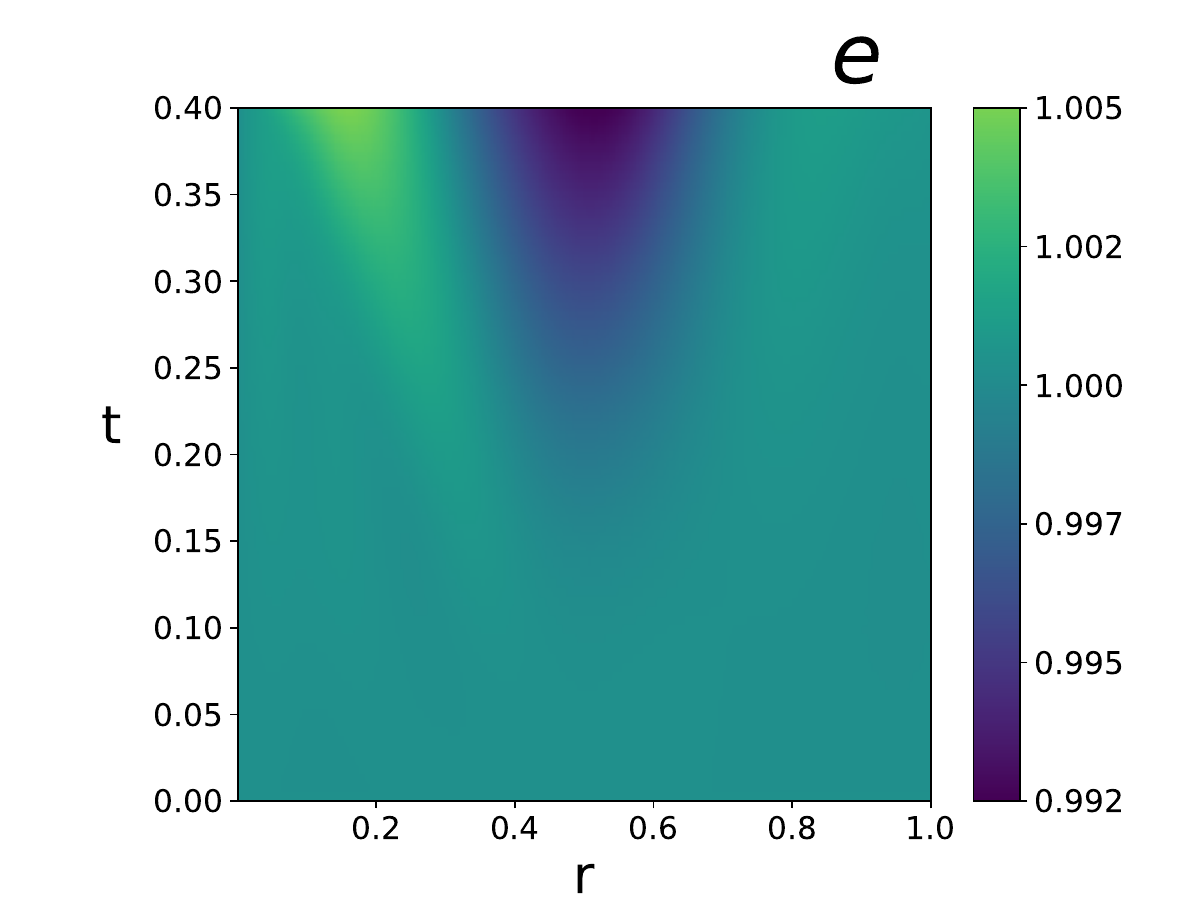}
\end{minipage}
\caption{
Heatmaps in the $t$-$r$ plane of the radial velocity $u^r$, the angular velocity $u^\varphi$,
and the thermodynamic energy density $e$, shown from left to right, for $\gamma=2$.
}
\label{Fig:ene_EdH}
\end{figure*}

\subsubsection{Impact on the flow profile}
\label{sec:Feedback-flow-EdH}

Finally, we discuss the evolution of the energy density and fluid flow induced by the rotational viscous effect.
In Fig.~\ref{Fig:ene_EdH}, we show the heatmaps in the $t$-$r$ plane of the radial velocity $u^r$, the angular velocity $u^\varphi$,
and the thermodynamic energy density $e$ as in the previous subsection.

We find that $u^r$ and $u^\varphi$ gradually develop from their vanishing initial conditions.
They are purely induced by the initial spin angular momentum through the rotational viscous effect.
The radial flow $u^r$ acquires both negative and positive peaks,
indicating emergence of an inward flow against the centrifugal force.
The two-peak structure in $u^\varphi$ further shows that the interior and exterior regions of the cylinder rotate in opposite directions, as sketched in Fig.~\ref{Fig:spin_to_orb}.
This position-dependent growth of the rotational flow can be understood from the evolution equation at the initial time $t = 0$,
$\frac{4}{3}e\partial_t (r u^{\varphi}) = -\partial_r \phi^{xy}$.
This indicates that $u^{\varphi}$ is generated by the radial derivative of the couple-stress tensor $\phi^{xy}$.
The thermodynamic energy density $e$ shows a larger magnitude near the center than in the exterior region.
This behavior again overrides the naive expectation based solely on the centrifugal force, and
provides further evidence of the rotational viscous effect.


\section{Summary}\label{Sec:S}

We have presented the first numerical analysis of relativistic spin
hydrodynamics with special care of the numerical treatment of the angular
momentum conservation using PINNs.  We have considered a rotating fluid confined in
a cylindrical container to investigate the conversions of spin and orbital
angular momentum, the Barnett and Einstein--de Haas effects, in the fully nonlinear
evolution of relativistic spin hydrodynamics for a finite time range.

In the first part of the paper, we have shown that PINNs can successfully solve the relativistic spin hydrodynamic equations
while accurately preserving the total angular momentum.
This indicates that our PINNs-based prescription offers an approach to addressing the challenge of angular momentum conservation
in computational relativistic spin hydrodynamics.

In the latter part of the paper, we have investigated the role of the rotational viscous effect
in driving the conversion between spin and orbital angular momentum.
Two complementary scenarios have been studied: a case where orbital angular momentum is converted into spin angular momentum,
and the opposite case where spin angular momentum is converted into orbital angular momentum.
This conversion effect has been confirmed in our numerical analysis on the time evolution of relativistic spinful fluid.
Furthermore, by analyzing the couple-stress tensor, the transverse thermal vorticity, and the spin potential,
we have revealed the detailed mechanism of the conversion process:
an initial imbalance between the transverse thermal vorticity and the spin potential enhances the couple-stress tensor,
which subsequently leads to the transfer between orbital and spin angular momentum.

It should be noted that this conversion process, particularly the transfer from orbital to spin angular momentum,
may play an important role in the spin generation in high-energy heavy-ion collisions.
Furthermore, for both simulation setups, the rotational viscous effect not only leads to the spin-orbit conversion
but also largely influences the time evolution of the hydrodynamic flow during the finite time range, which is nonlinear in nature.
This underscores the importance of fully nonlinear relativistic spin hydrodynamics. 

No numerical simulations of fully nonlinear relativistic spin hydrodynamics have been performed to date,
and their nonequilibrium and nonlinear properties remain largely unexplored.
Against this background, this study lays a basis for future accurate numerical studies of relativistic spin hydrodynamics based on PINNs.
In particular, comparisons of nonlinear numerical results with linear analyses will be essential for understanding nonlinear effects in relativistic spin hydrodynamics,
which represent a first step toward the further development of the theory.
It is also of interest to examine how the system behaves when the spin potential and thermal vorticity are treated at the zeroth-order,
or when second-order effects, such as those induced by the boost heat current, are incorporated.

The method is also important in developing dynamical models used for the
quantitative description of high-energy heavy-ion collision reactions.
Although the direct use of PINNs for event-by-event simulations is challenged
by the large training cost of PINNs, the PINNs solutions can be used as
reference in calibrating and validating relativistic spin hydrodynamic codes
based on conventional numerical schemes.  
Finally, we conclude that approaches based on PINNs open possibilities for general physics problems
where specific physical constraints are important but difficult to enforce numerically in conventional methods.


\section*{Acknowledgments}
We acknowledge that numerical calculations were performed on Supercomputer
Yukawa-21 at Yukawa Institute for Theoretical Physics (YITP) at Kyoto
University.
This work is supported by the JSPS KAKENHI under Grant Nos.~JP22H01216,
JP23H05439, and~JP23K13102 and by the National Natural Science Foundation of China (NSFC) under grant numbers W2433010 and W2532002.


\section*{Data availability}
The data and scripts necessary to reproduce the main numerical results
and figures are available in Ref.~\cite{MatsudaData2026}.

\appendix

\section{Spin Susceptibility}\label{App:spin_sus}
We derive the spin susceptibility for a system composed of massless two-flavor fermions and gluons in the rarefied gas limit.
The spin susceptibility characterizes how sensitively the spin is induced in response to perturbations on the spin potential.
We focus on the linear regime with an infinitesimal perturbation.
The expression for the spin susceptibility we use is given as
\begin{align}
\chi &= \frac{\partial \average{ S^{\mu\nu} }}{\partial \omega_{\mu \nu}}
= \frac{2V}{T} \average{ (S^{\mu\nu})^2 }\ ,
\end{align}
where we assume rotational symmetry and $\langle S^{\mu\nu} \rangle = 0$,
and, for the purposes of this section,
we focus on the case where both the spin potential and chemical potential vanish
in an equilibrium state.
The brackets $\langle \cdot \rangle$ denote thermal averages in this system.
Without loss of generality, we consider the $\mu=x$ and $\nu=y$ components.
For an ideal gas, where the gauge and fermion fields are not coupled, the susceptibility is given as the sum of the susceptibilities for each field,
\begin{align}
\chi = \chi_{\rm G} + \chi_{\rm F}\ .
\end{align}
The spin operators for the gauge and fermion fields are linear combinations of the number operators as
\begin{align}
S^{xy}_{\rm G} &= N^\uparrow_{\rm G} - N^\downarrow_{\rm G}\ , \\
S^{xy}_{\rm F} &= \frac{1}{2}\left[N^\uparrow_{\rm F} - N^\downarrow_{\rm F}\right]\ ,
\end{align}
where $\uparrow$ and $\downarrow$ represent spin orientations along the $z$-axis,
corresponding to the positive and negative directions of the axis, respectively,
and fermion-antifermion symmetry is assumed, meaning that the expectation values of fermion and antifermion numbers are equal.
Thus, the spin susceptibilities are related to the number susceptibility as follows
\begin{align}
\chi_{\rm G} &= \frac{2V}{T} \left[ \average{ (N^\uparrow_{\rm G})^2 } + \average{ (N^\downarrow_{\rm G})^2 } 
\right] = 32 \chi_{N, \rm B}\ ,\\
\chi_{\rm F} &= \frac{V}{2T} \left[ \average{ (N^\uparrow_{\rm F})^2 } + \average{ (N^\downarrow_{\rm F})^2 }
\right] = 12 \chi_{N, \rm F}\ ,
\end{align}
where $\chi_{N, \rm B/F}$ is the number susceptibility for a single degree of freedom,
and it can be calculated using the Bose/Fermi distribution,
\begin{align}
\chi_{N, \rm B/F} &= \int \frac{d^3k}{(2\pi)^3} \left.\frac{\partial f_{\rm B/F}(\omega)}{\partial \mu} \right|_{\mu=0} \nonumber\\
                  &= \int \frac{d^3k}{(2\pi)^3} \beta f_{\rm B/F}(\omega) \left[ 1 \pm f_{\rm B/F}(\omega) \right]\Bigr|_{\mu=0}\ .
\end{align}
In the massless limit, these integrals yield
\begin{align}
\chi_{N, \rm B} = \frac{T^2}{6}\ ,\ \ \ \chi_{N, \rm F} = \frac{T^2}{12}\ .
\end{align}
In summary, the spin susceptibility is obtained as
\begin{align}
\chi = \frac{19}{3} T^2 = \frac{19}{3} \left( \frac{30 e}{37\pi^2} \right)^{1/2}\ .
\end{align}
In the last line of this equation, we use the expression for the energy in the current setup, $e = \frac{37}{30}\pi^2 T^4$.

\section{Hydrodynamic Equations}\label{App:hydro_eq}
Here, we present the detailed expressions for the hydrodynamic equations that we are working with,
as derived in Sec.~\ref{Sec:2_disk}.
First, Eq.~\eqref{Eq:hydro1} is given by
\begin{align}
&u^\mu \partial_\mu e = -\frac{4}{3}e \left( \partial_\mu u^\mu + \frac{1}{r} u^r \right)
- 2r u^\varphi \phi^{r \varphi} \nonumber \\
&- \frac{r^2 \phi^{r\varphi}}{u^t} \left( u^t \partial_r u^\varphi - u^\varphi \partial_r u^t + u^r \partial_t u^\varphi - u^\varphi \partial_t u^r \right)
\ .
\end{align}
Next, Eqs.~\eqref{Eq:hydro2} and \eqref{Eq:hydro3} are
\begin{align}
&\frac{4}{3}e u^\mu \partial_\mu  u^r
= \frac{4}{3} r e (u^\varphi)^2 - \frac{1}{3} u^r u^\mu \partial_\mu  e - \frac{1}{3}\partial_r e
+ 2r u^r u^\varphi \phi^{r \varphi} \nonumber \\
& + \frac{r^2 u^r \phi^{r \varphi}}{u^t} \left( u^t \partial_r u^\varphi - u^\varphi \partial_r u^t + u^r \partial_t u^\varphi - u^\varphi \partial_t u^r \right) \nonumber \\
& + \frac{r^2}{(u^t)^2} \left[ u^t (\partial_t u^\varphi) \phi^{r \varphi} - u^\varphi (\partial_t u^t) \phi^{r \varphi} + u^t u^\varphi \partial_t \phi^{r \varphi} \right]\ ,\\
&\frac{4}{3}e u^\mu \partial_\mu  u^\varphi
= -\frac{8}{3r}e u^r u^\varphi - \frac{1}{3}u^\varphi u^\mu \partial_\mu  e - \frac{1}{r} \phi^{r \varphi}\nonumber\\
&+ 2r (u^\varphi)^2 \phi^{r \varphi}
-\partial_r \phi^{r \varphi}
\nonumber \\
& + \frac{r^2 u^\varphi \phi^{r \varphi}}{u^t} \left( u^t \partial_r u^\varphi - u^\varphi \partial_r u^t + u^r \partial_t u^\varphi - u^\varphi \partial_t u^r \right)\nonumber \\
& - \frac{1}{(u^t)^2} \left[u^t (\partial_t u^r) \phi^{r \varphi} - u^r (\partial_t u^t) \phi^{r \varphi}
+ u^t u^r \partial_t \phi^{r \varphi} \right]
\ .
\end{align}
There is no further need to expand Eq.~\eqref{Eq:hydro4}.
Moving on to Eq.~\eqref{Eq:hydro5}, we have
\begin{align}
&\tau_\phi u^\mu \partial_\mu  \phi^{r \varphi}
= - \phi^{r \varphi} \nonumber \\
&+ \tau_\phi \phi^{r\varphi}
\left[ \frac{1}{u^t} u^\mu \partial_\mu  u^t - \frac{2}{3} \partial_\mu u^\mu - \frac{5}{3r} u^r \right] \nonumber \\
&- \gamma \left[ u^r u^\mu \partial_\mu  u^\varphi - u^\varphi u^\mu \partial_\mu  u^r + \partial_r u^\varphi + r (u^\varphi)^3 \right] \nonumber \\
&-\frac{2\gamma}{r}\left\{\left[1+(u^r)^2\right]u^\varphi - \frac{6 \pi }{19} \left( \frac{37}{30} \right)^{\frac{1}{2}}\frac{u^t S^z}{\sqrt{e}}\right\}\ .
\end{align}

\section{Boundary Conditions}\label{App:BC}
We consider a cylindrical system with radius $R$,
and impose boundary conditions at both $r = 0$ and $r = R$ to ensure the uniqueness of the solution.
First, we discuss the boundary at $r=0$.
The hydrodynamic equations shown in Appendix~\ref{App:hydro_eq} contain terms proportional to $1/r$,
which can diverge at $r \to 0$ unless appropriate boundary conditions are applied.
To eliminate such divergences, we impose the following Dirichlet boundary conditions,
\begin{align}
e(t,r=0)              &= e(t=0,r=0)\ ,\\
u^r(t,r=0)            &= 0\ ,\\
u^\varphi(t,r=0)       &= u^\varphi(t=0,r=0)      \ ,\\
S^z(t,r=0)            &= S^z(t=0,r=0)      \ ,\\
\phi^{r\varphi}(t,r=0) &= 0\ ,
\end{align}
with
\begin{align}
u^\varphi(t,r=0)
-
\frac{6\pi}{19} \left(\frac{37}{30}\right)^{\frac{1}{2}}
\frac{S^z(t,r=0)}{\sqrt{e(t,r=0)}}= 0\ .
\end{align}
To avoid this divergence, the coefficient attached to $1/r$ must approach zero at $r=0$ at a rate equal to or faster than $r$.
Therefore, this boundary condition alone is insufficient to prevent divergence throughout the entire time evolution.
However, since PINNs can inherently represent only smooth functions,
they may automatically approximate a seemingly regular solution around $r=0$,
even if the true solution exhibits singular behavior there.

Next, we consider the boundary condition at $ r = R $.
We intend to impose the boundary condition at $ r = R $ such that energy and angular momentum are conserved within the cylinder.
The continuity equations for energy and angular momentum, $\nabla_\mu T^{\mu t} = 0 $ and $ \nabla_\mu J^{\mu xy} = 0 $, give
\begin{align}
\partial_\mu (r T^{\mu t} ) &= 0\ , \\
\partial_\mu (r J^{\mu xy}) &= 0\ ,
\end{align}
where the derivative $\partial_\mu$ is taken in cylindrical coordinates.
This leads to the conservation laws,
\begin{align}
\partial_t \left( \int_0^R dr\, r\, T^{tt} \right) &= - \int_0^R dr \, \partial_r ( r T^{rt}) \nonumber \\
&= - R \left.\left( \frac{4}{3} e\, u^t u^r - \frac{R\, u^\varphi\, \phi^{r\varphi}}{u^t} \right)\right|_{r=R} \ , \\
\partial_t \left( \int_0^R dr\, r\, J^{txy} \right) &= - \int_0^R dr \, \partial_r ( r J^{rxy} ) \nonumber \\
&= - R^3 \left.\left( \frac{4}{3} e\, u^r u^\varphi + \phi^{r\varphi} \right)\right|_{r=R} \ .
\end{align}
These quantities are conserved if we impose the following Dirichlet-type conditions,
\begin{align}
u^r(t,r=R) = \phi^{r\varphi}(t,r=R) = 0\ .
\end{align}
Furthermore, the hydrodynamic equation for the spin density, $ \partial_t S^z = -2r \phi^{r\varphi} $,
implies that the spin density must also be fixed at $ r = R $:
\begin{align}
S^z(t,r=R) = S^z(t=0,r=R)\ .
\end{align}
The remaining boundary conditions at $ r = R $ are given in the Neumann type,
meaning that the radial derivatives of the variables are kept constant over time.
Specifically, we impose
\begin{align}
\partial_r u^r(t,r=R) = \partial_r \phi^{r\varphi}(t,r=R) = 0\ ,
\end{align}
but we do not impose the Neumann conditions on $e$ and $u^\varphi$ at $r = R$.

\section{Improvement of angular momentum conservation}\label{App:comp}
We verify that including the penalty term for the angular momentum conservation law, i.e., the last term in Eq.~\eqref{Eq:loss}, successfully suppresses its violation
by comparing the numerical results with and without the penalty term.
The numerical setup used in this section is as follows.
The number of hidden layers is two.
The batch size is set to $2N_{\text{col}} + N_{\text{col,global}} = 2N_{\text{col}} + N_t N_r = 2 \times 25,000 + 5 \times 1,000 = 55,000$.
All other numerical parameters are the same as those presented in the main part of the paper.
The initial conditions are given in Eqs.~\eqref{Eq:init_orb_1}--\eqref{Eq:init_orb_5} using the parameters summarized in Table~\ref{Tab:para_app}\@.
These parameters, as well as the physical quantities discussed below, are normalized with respect to $\bar{e}$.
We set the full time domain up to $t_{\text{max}} = 0.2$ from the start, without updating the time domain during training.
The numerical results shown in this section are obtained after 100,000 iterations. 
The training takes approximately 1 hour, using one NVIDIA H100 NVL GPU.

\begin{table}[htbp]
\begin{center}
\begin{tabular}{r|c}
\hline\hline
$\bar{e}$      & $1$ \\
$R$            & $1$ \\
$t_\text{max}$ & $0.2$ \\
$\gamma_\phi$  & $2$ \\
$\tau_\phi$    & $2$ \\
$\delta_1$     & $0.2$ \\
\hline\hline
\end{tabular}
\end{center}
\caption{
Parameters used in Appendix~\ref{App:comp}, expressed in units of $\bar{e}$.
}
\label{Tab:para_app}
\end{table}
\begin{figure*}[tp]
\begin{minipage}{0.4\textwidth}
    \centering
    \includegraphics[width=\textwidth]{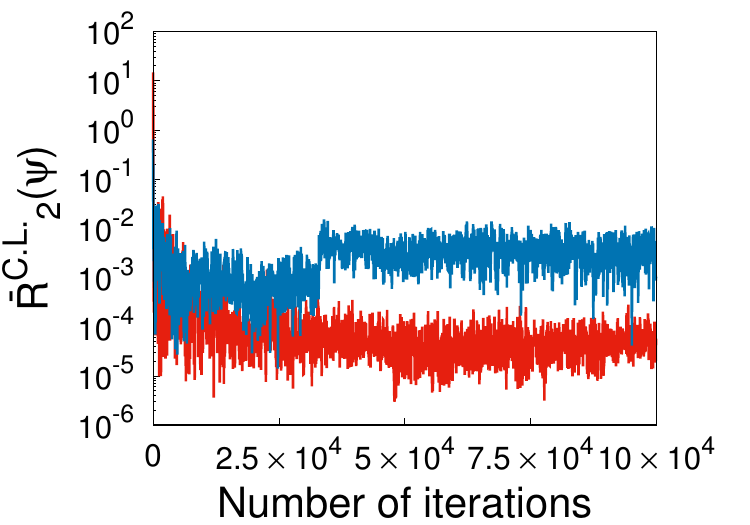}
\end{minipage}
\caption{
Decrease in the averaged residual associated with the global conservation of angular momentum,
as defined in Eq.~\eqref{Eq:residual_global}.
The red line corresponds to the case with the penalty term enforcing angular momentum conservation
while the blue line corresponds to the case without the penalty term.
}
\label{Fig:comp}
\end{figure*}
\begin{figure*}[tp]
\begin{minipage}{0.4\textwidth}
    \centering
    \includegraphics[width=\textwidth]{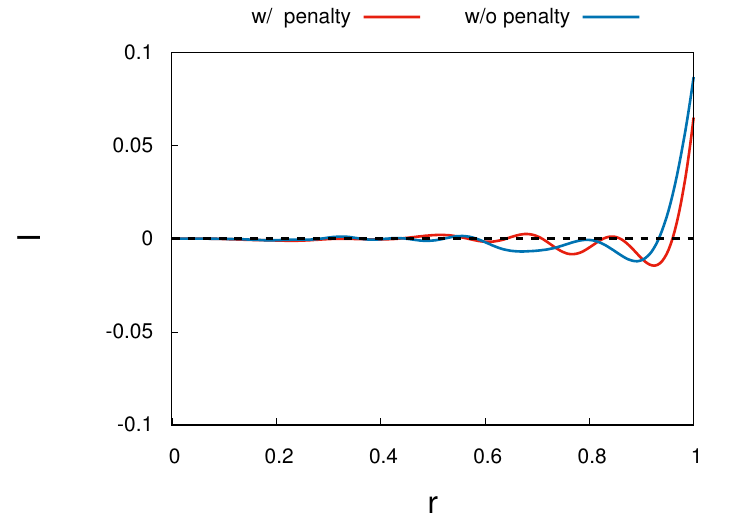}
\end{minipage}
\begin{minipage}{0.4\textwidth}
    \centering
    \includegraphics[width=\textwidth]{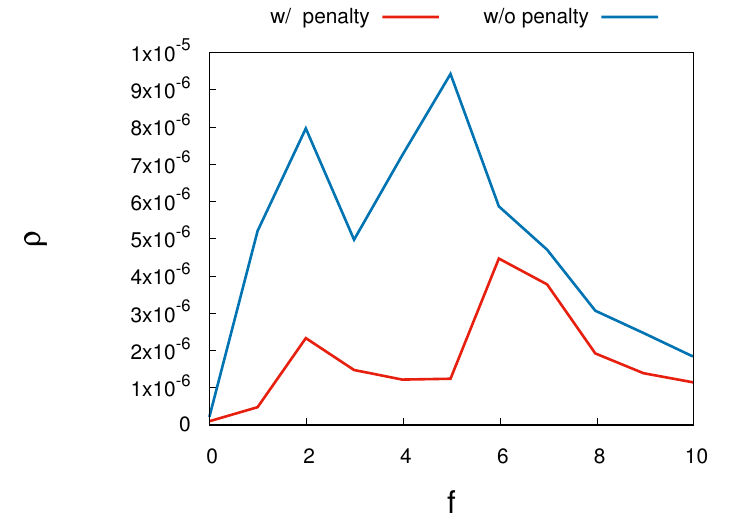}
\end{minipage}
\caption{
Left: Radial distribution of $I(\psi; r)$,
the time-integrated local violation of angular momentum conservation defined in Eq.~\eqref{Eq:I}.
Right: Power spectrum of $I(\psi; r)$ defined in Eq.~\eqref{Eq:Rho}.
}
\label{Fig:ang_Metho}
\end{figure*}

In Fig.~\ref{Fig:comp},
we show the decrease in the averaged residual associated with the global conservation of angular momentum,
as defined in Eq.~\eqref{Eq:residual_global},
during the training processes described above.
The residual decreases to the range from $10^{-4}$ to $10^{-6}$ when the penalty is imposed.
Without the penalty term, the residual still decreases during training, reaching the range from $10^{-2}$ to $10^{-4}$; however, it remains larger than in the case with the penalty term. In addition, its convergence behavior is found to be less stable. This indicates that the penalty term is essential for enforcing global angular momentum conservation with high accuracy.

We then consider the local violation of angular momentum conservation, as quantified by $r R_\text{local}(\psi; t, r)$,
after $10^5$ training iterations in the same simulations for Fig.~\ref{Fig:comp}.
The integration of this local violation is directly related to the violation of the global conservation of angular momentum.
In the left panel of Fig.~\ref{Fig:ang_Metho}, we show the radial distribution of $r R_\text{local}$ integrated over $t$, defined as
\begin{align}
I(\psi; r) \equiv \frac{1}{t_\text{max}} \int^{t_\text{max}}_0 dt \, r R_\text{local}(\psi; t, r)\ ,\label{Eq:I}
\end{align}
where it is normalized by $t_\text{max}$.
The results show that the penalty term reduces the overall magnitude of the local violation.
It is also noted that a positive amplification of the violation is observed near $r = R$.

In the right panel of Fig.~\ref{Fig:ang_Metho}, we show the power spectrum of $I(\psi; r)$ defined as
\begin{align}
\rho(f) = \left| \int^R_0 dr \, I(r) e^{- 2i\pi f r} \right|^2\ .\label{Eq:Rho}
\end{align}
The numerical result shows that the penalty term reduces the violation of angular momentum conservation particularly in the low-frequency regime, $f \leq 5$.
This reduction in the low-frequency regime suggests that an impact of violation on the long-wavelength dynamics of the fluid is negligible.

\bibliography{main}

\end{document}